\documentclass[pdflatex,sn-mathphys]{sn-jnl}

\jyear{2021}%

\theoremstyle{thmstyleone}%
\theoremstyle{thmstyletwo}%

\theoremstyle{thmstylethree}%

\usepackage{acro}
\DeclareAcronym{td}{short=TD, long=Technical Debt, tag=abbrev}
\DeclareAcronym{satd}{short=SATD, long=Self-Admitted Technical Debt, tag=abbrev}
\DeclareAcronym{cvmtd}{short=CVM-TD, long=Contextualized Vocabulary Model, tag=abbrev}
\DeclareAcronym{nlp}{short=NLP, long=Natural Language Processing, tag=abbrev}
\DeclareAcronym{ast}{short=AST, long=Abstract Syntax Tree, tag=abbrev}
\DeclareAcronym{sbt}{short=SBT, long=Structure-Based Traversal, tag=abbrev}
\DeclareAcronym{oom}{short=OOM, long=Out Of Memory, tag=abbrev}
\DeclareAcronym{rnn}{short=RNN, long=Recurrent Neural Network, tag=abbrev}
\DeclareAcronym{lstm}{short=LSTM, long=Long Short-Term Memory, tag=abbrev}
\DeclareAcronym{ffnn}{short=FFNNs, long=Feed-Forward Neural Networks, tag=abbrev}
\DeclareAcronym{cv}{short=CV, long=Cross Validation, tag=abbrev}
\DeclareAcronym{nmt}{short=NMT, long=Neural Machine Translation, tag=abbrev}
\DeclareAcronym{tp}{short=TP, long=True Positive, tag=abbrev}
\DeclareAcronym{tn}{short=TN, long=True Negative, tag=abbrev}
\DeclareAcronym{fp}{short=FP, long=False Positive, tag=abbrev}
\DeclareAcronym{fn}{short=FN, long=False Negative, tag=abbrev}
\DeclareAcronym{p}{short=P, long=Precision, tag=abbrev}
\DeclareAcronym{r}{short=R, long=Recall, tag=abbrev}
\DeclareAcronym{mnb}{short=MNB, long=Multinomial Naive Bayes, tag=abbrev}
\DeclareAcronym{svm}{short=SVM, long=Support Vector Machines, tag=abbrev}
\DeclareAcronym{bow}{short=BoW, long=Bag of Words, tag=abbrev}
\DeclareAcronym{tfidf}{short=TF-IDF, long=Term Frequency-Inverse Document Frequency, tag=abbrev}
\DeclareAcronym{gru}{short=GRU, long=Gated Recurrent Unit, tag=abbrev}
\DeclareAcronym{ml}{short=ML, long=Machine Learning, tag=abbrev}
\DeclareAcronym{dl}{short=DL, long=Deep Learning, tag=abbrev}
\DeclareAcronym{cpl}{short=CPL, long=Comment Processing Line, tag=abbrev}
\DeclareAcronym{scpl}{short=SCPL, long=Source Code Processing Line, tag=abbrev}
\DeclareAcronym{vsm}{short=VSM, long=Vector Space Model, tag=abbrev}
\DeclareAcronym{cnn}{short=CNN, long=Convolutional Neural Network, tag=abbrev}
\DeclareAcronym{tml}{short=TML, long=Traditional Machine Learning, tag=abbrev}
\DeclareAcronym{rf}{short=RF, long=Random Forest, tag=abbrev}
\DeclareAcronym{sat}{short=SATs, long=Static Analysis Tools, tag=abbrev}
\DeclareAcronym{oov}{short=OOV, long=Out of Vocabulary, tag=abbrev}

\begin{document}

\title[Technical Debt Identification and Description]{A Framework for Conditional Statement Technical Debt Identification and Description}


\author*[1]{\fnm{Abdulaziz} \sur{Alhefdhi}}\email{aa043@uowmail.edu.au}

\author[1]{\fnm{Hoa} \sur{Khanh~Dam}}\email{hoa@uow.edu.au}

\author[2]{\fnm{Yusuf} \sur{Sulistyo~Nugroho}}\email{yusuf.nugroho@ums.ac.id}

\author[3]{\fnm{Hideaki} \sur{Hata}}\email{hata@shinshu-u.ac.jp}

\author[4]{\fnm{Takashi} \sur{Ishio}}\email{ishio@is.naist.jp}

\author[1]{\fnm{Aditya} \sur{Ghose}}\email{aditya@uow.edu.au}

\affil*[1]{\orgdiv{School of Computing and Information Technology}, \orgname{University of Wollongong}, \orgaddress{\city{Wollongong}, \postcode{2522}, \state{NSW}, \country{Australia}}}

\affil[2]{\orgdiv{Teknik Informatika}, \orgname{Universitas Muhammadiyah Surakarta}, \orgaddress{\city{Surakarta}, \postcode{57102}, \state{Jawa Tengah}, \country{Indonesia}}}

\affil[3]{\orgdiv{Faculty of Engineering}, \orgname{Shinshu University}, \orgaddress{\city{Wakasato}, \postcode{380-8553}, \state{Nagano}, \country{Japan}}}

\affil[4]{\orgdiv{Division of Information Science}, \orgname{Nara Institute of Science and Technology}, \orgaddress{\city{Ikoma}, \postcode{630-0192}, \state{Nara}, \country{Japan}}}


\abstract{\ac{td} occurs when development teams favour short-term operability over long-term stability. Since this places software maintainability at risk, technical debt requires early attention to avoid paying for accumulated interest. Most of the existing work focuses on detecting technical debt using code comments, known as Self-Admitted Technical Debt (SATD). However, there are many cases where technical debt instances are not explicitly acknowledged but deeply hidden in the code. In this paper, we propose a framework that caters for the absence of SATD comments in code. Our Self-Admitted Technical Debt Identification and Description (\emph{SATDID}) framework determines if technical debt should be self-admitted for an input code fragment. If that is the case, SATDID will automatically generate the appropriate descriptive SATD comment that can be attached with the code. While our approach is applicable in principle to any type of code fragments, we focus in this study on technical debt hidden in conditional statements, one of the most TD-carrying parts of code. We explore and evaluate different implementations of SATDID. The evaluation results demonstrate the applicability and effectiveness of our framework over multiple benchmarks. Comparing with the results from the benchmarks, our approach provides at least 21.35\%, 59.36\%, 31.78\%, and 583.33\% improvements in terms of Precision, Recall, F-1, and Bleu-4 scores, respectively. In addition, we conduct a human evaluation to the SATD comments generated by SATDID. In 1-5 and 0-5 scales for Acceptability and Understandability, the total means achieved by our approach are 3.128 and 3.172, respectively.}




\maketitle

\section{Introduction}\label{sect:intro}

Development teams aim at implementing software projects of high quality, on time, and within budget. In real-world scenarios however, quality sometimes is traded in order to deliver the software on time. Developers tend to apply quick fixes or temporary implementations that are not necessarily ideal for the long term, which can lead to the incurring of technical debt \cite{brown2010managing,tom2013exploration}.

Technical Debt (TD) is a term introduced by Ward Cunningham \cite{cunningham1993wycash} to describe the situation where accomplishing short-term goals is chosen over long-term code quality. Just like financial debt, \ac{td} can accumulate interest if it is not dealt with quickly. Poor coding practices can cause the presence of \ac{td}. Previous studies have highlighted the widespread occurrence of \ac{td} and its impact on software quality, complexity, and maintenance. The presence of \ac{td} makes changes to the system more frequent \cite{zazworka2011investigating} and harder to implement \cite{wehaibi2016examining}. Developers recognise that \ac{td} is unavoidable and, therefore, in need of careful management \cite{lim2012balancing}. Repaying \ac{td} comes in the form of re-structuring and refactoring the software \cite{huang2018identifying}.

Despite the importance of \ac{td} management, especially in its early stages, the identification of \ac{td} in itself is a challenge. Moreover, \ac{td} needs to be well-understood once it is identified in order to be properly managed. For example, if a developer implements a \emph{workaround} in the code and wraps it within a \emph{conditional statement} (e.g. \texttt{if}-statement), other developers in the team may not know that this is a \ac{td}-carrying statement. To make it visible, the developer writes a comment that describes the \ac{td}. Such comments flag \ac{satd} \cite{potdar2014exploratory}.

The majority of existing work \cite{potdar2014exploratory,maldonado2015detecting,de2015contextualized,da2017using,huang2018identifying,yan2018automating,8661216,Maipradit2020,9252045} has focused on developing tool support for detecting if a code \underline{comment} flags \ac{satd}. They presume the existence of such comments attached to code fragments that contain technical debt. \emph{However, this presumption does not always hold in practice. There are many instances where \ac{td} in code is not explicitly acknowledged in the form of a comment}. In these cases, there is a need for automated machinery which can: (i) determine if a given code fragment introduces technical debt; and if so, (ii) generate the appropriate (self-admitting) comment that can be attached to the code. However, proposals to provide this kind of support is currently missing in the literature. 

In this paper, we propose a framework that provides two levels of support: \ac{satd} recommendation and \ac{satd} description. Given a code fragment as input, our Self-Admitted Technical Debt Identification and Description framework (\emph{SATDID}) determines/identifies if technical debt should be self-admitted for this code fragment (level 1), and then automatically generates a comment admitting and describing the detected technical debt instance (level 2). SATDID can be used on-the-fly to recommend and describe potential \ac{satd} as the developers writes the code. This way, \ac{td} can be prevented before its occurrence. We explore and evaluate the capabilities of different machine/deep learning approaches in implementing SATDID and providing these levels of support.

Although our approach can generally be applied to any size and type of code fragments, our focus in this study is on conditional statements. Conditionals are prominent in the context of technical debts. Kruchten et al. \cite{kruchten2019managing} place ``quick-and-dirty" conditional statements on the top of their \ac{td} example list for managing technical debt. In addition, previous studies (e.g. \cite{zampetti2020automatically,zampetti2018self}) found that \ac{satd} comments were often associated with conditional statements.

Our contribution in this paper is as follows:

\begin{enumerate}
    \item \textbf{A dual-layered framework for SATD management:} To the best of our knowledge, SATDID is the first to provide two levels of support for SATD management, i.e. SATD recommendation and SATD comment generation.
    
    \item \textbf{Leveraging machine/deep learning for SATD management:} SATDID consists of different machine/deep learning components that are carefully developed and improves the results over all the baselines.
    
    \item We build and publish the first \textbf{dataset} of labelled SATD and non-SATD code-comment pairs consisting of conditional statements and their associated comments. We also made our \textbf{code-base} and experiment reports publicly available\footnote{\url{https://github.com/Abdulaziz-Alhefdhi/tech_debt}}.
\end{enumerate}

The rest of the paper is outlined as follows. Section~\ref{sect:motivation} gives a motivating example. Section~\ref{sect:design} illustrates the architecture of our framework (SATDID). The paper in the following sections describes the key components of SATDID in details.
Section~\ref{sect:process-vector} describes the first two modules, namely Data Processing and Data Vectorisation. Section~\ref{satd-id} describes the SATD Identification module. Section~\ref{satd-com-gen} describes the SATD Comment Generation module. After that, we explain the model training processes in Section~\ref{trn-dl}. We evaluate and discuss our approach in Sections~\ref{sect:evaluation} and \ref{sect:discuss}. We present the related work in Section \ref{sect:related-work} before we conclude our study in Section~\ref{sect:conclusion}. 
\section{Motivating example}\label{sect:motivation}

Previous research \cite{wehaibi2016examining} studied the impact of Technical Debt (\ac{td}) on software complexity and changeability. \ac{td} needs to be addressed, and the ultimate goal is to remove (repay) \ac{td} instances from the software. In order to manage and eventually repay \ac{td}, we first need to identify its occurrences in the codebase. This is a challenging task, especially in large software projects where codebases can grow to millions of lines of code. Figure \ref{motiv_eg} depicts several scenarios to illustrate various challenges of this problem. 




\begin{figure}[htb]

    \fcolorbox{black}{green!25}{Scenario 1.a}
    
    \scriptsize
    
    \hfill
    
    \texttt{\textcolor{gray}{// make sure someone didn't whack the clockSeqAndNode by}}
    
    \texttt{\textcolor{gray}{// changing the order of instantiation.}}
    
    \texttt{\textcolor{BlueViolet}{if (}clockSeqAndNode \textcolor{BlueViolet}{==} \textcolor{OliveGreen}{0}\textcolor{BlueViolet}{)}}
    
    \qquad \texttt{\textcolor{BlueViolet}{throw new} RuntimeException\textcolor{BlueViolet}{(}\textcolor{OliveGreen}{"singleton instantiation is}}
    
    \qquad \qquad \qquad \qquad \qquad \qquad \qquad \qquad ~\texttt{\textcolor{OliveGreen}{misplaced."}\textcolor{BlueViolet}{);}}
    
    \hfill
    
    \hfill
    
    \normalsize
    
    \fcolorbox{black}{yellow!25}{Scenario 1.b}
    
    \scriptsize
    
    \hfill
    
    \texttt{\textcolor{gray}{// @todo: I'm not sure we need an error here}}
    
    \texttt{\textcolor{BlueViolet}{if (!}ade\textcolor{BlueViolet}{.}versionOnly\textcolor{BlueViolet}{)}}
    
    \qquad \texttt{ade\textcolor{BlueViolet}{.}de\textcolor{BlueViolet}{.}getDiskId\textcolor{BlueViolet}{().}setPendingAsync\textcolor{BlueViolet}{(}\textcolor{RedOrange}{false}\textcolor{BlueViolet}{);}}
    
    \hfill
    
    \hfill
    
    \normalsize
    
    \fcolorbox{black}{green!25}{Scenario 2.a}
    
    \scriptsize
    
    \hfill
    
    \texttt{\textcolor{BlueViolet}{if (}log\textcolor{BlueViolet}{.}isTraceEnabled\textcolor{BlueViolet}{()) \{}}
    
    \qquad \texttt{log\textcolor{BlueViolet}{.}trace\textcolor{BlueViolet}{(}\textcolor{OliveGreen}{"Searching for: \{\} in package: \{\} using}}
    
    \qquad \qquad \qquad \texttt{\textcolor{OliveGreen}{classloader: \{\}"}\textcolor{BlueViolet}{, new} Object\textcolor{BlueViolet}{[]\{}test\textcolor{BlueViolet}{,} packageName\textcolor{BlueViolet}{,}}
    
    \qquad \qquad \qquad \texttt{loader\textcolor{BlueViolet}{.}getClass\textcolor{BlueViolet}{().}getName\textcolor{BlueViolet}{()\});}}
	
	\texttt{\textcolor{BlueViolet}{\}}}
    
    \hfill
    
    \hfill
    
    \normalsize
    
    \fcolorbox{black}{red!25}{Scenario 2.b}
    
    \scriptsize
    
    \hfill
    
    \texttt{\textcolor{BlueViolet}{if (!}getExpression\textcolor{BlueViolet}{().}isValid\textcolor{BlueViolet}{() \&\& !}forceRevalidated) \textcolor{BlueViolet}{\{}}
    
    \qquad \texttt{forceRevalidated \textcolor{BlueViolet}{=} \textcolor{RedOrange}{true}\textcolor{BlueViolet}{;}}
    
    \qquad \texttt{getExpression\textcolor{BlueViolet}{().}forceRevalidate\textcolor{BlueViolet}{();}}
    
    \texttt{\textcolor{BlueViolet}{\}}}

    
    \caption{In both Scenarios 1.a and 1.b, the conditional statement is accompanied with a comment while there is no comment accompanying the conditional statement in Scenarios 2.a and 2.b. The conditional statement in Scenarios 1.a and 2.a (highlighted with light green) is TD-free. The conditional statement in Scenario 1.b (highlighted with light yellow) contains \ac{td} which is noted by the preceding \ac{satd} comment. The conditional statement in Scenario 2.b (highlighted with light red) also contains \ac{td} but a \ac{satd} comment is missing.}
    
	\label{motiv_eg}
\end{figure}


Scenario 1.a in Figure \ref{motiv_eg} depicts the case of a \ac{td}-free conditional statement with an associated comment that describes what the \texttt{if}-statement does. However, in Scenario 1.b, there is \ac{td} in the conditional statement, which is \emph{explicitly} acknowledged in the associated comment that suggests code revision.
The challenge in Scenarios 1.a and 1.b is to detect which comment is a \ac{satd} comment and which one is not. Existing work in SATD (e.g. \cite{potdar2014exploratory,da2017using,huang2018identifying}) focuses on addressing this challenge.

The existing approaches rely on the comments provided with the code to detect \ac{satd}. However, there are many cases where technical debt is \emph{not} explicitly admitted in the form of a comment. Scenarios 2.a and 2.b depicts conditional statements that are not accompanied with comments. While the code in Scenario 2.a does not contain \ac{td}, the one in Scenario 2.b does. 

Scenario 2.b shows a conditional statements extracted from the infrastructure of project Openflexo\footnote{\url{https://support.openflexo.org/}}. In short, the \texttt{if}-statement was written to forcibly re-validate invalid \texttt{expressions} in the project. This is a temporary workaround. Ideally, the code should be written in a way that assures only valid expression production rather than let it potentially produce invalid expressions and re-validate them.
It is important to note that this \ac{td} is not self-admitted in a comment.

Automated support is thus needed to assist software engineers in scenarios similar to 2.a and 2.b. If a comment is not provided, the automated support analyses the source code to determine if it contains \ac{td} that should be self-admitted, and brings this to the software developer's attention. Upon receiving the developer's confirmation, the automated support generates an appropriate \ac{satd} comment and attach it with the code fragment. When used on-the-fly, the automated support raises a warning when a \ac{satd} comment is needed so that the developer decides to either modify the code or let the tool generate the appropriate \ac{satd} comment. In Section \ref{sect:design}, we illustrate the architectural design of our proposed framework that provides this kind of automated support.

\begin{figure}[htb]
	\includegraphics[width=\linewidth]{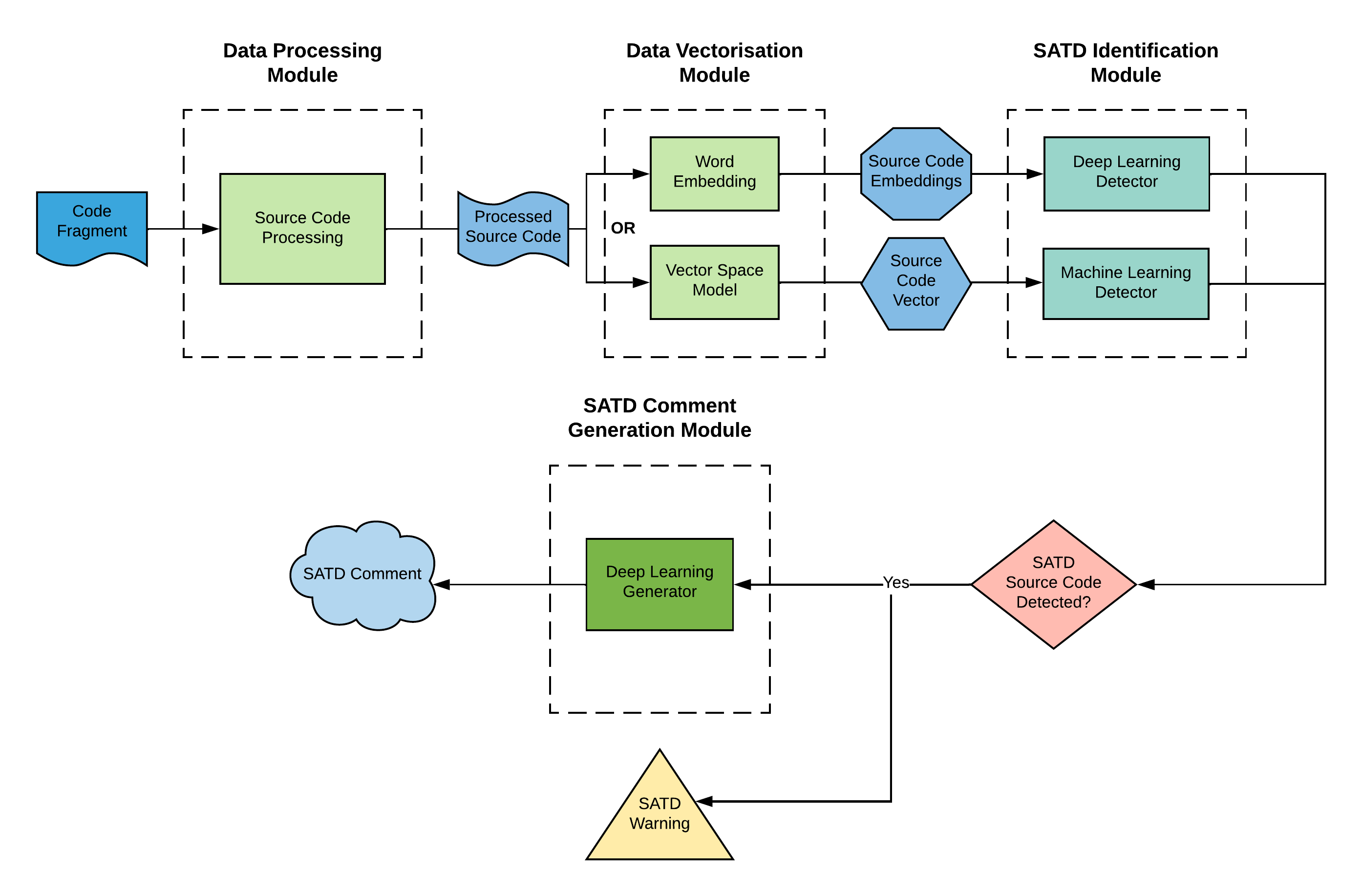}
	\caption{SATDID Architectural Design.}
	\label{fw_arch_des}
\end{figure}

\section{Architectural Design}\label{sect:design}



We propose an automated framework called SATDID that addresses the problems introduced by scenarios like the ones in Section \ref{sect:motivation}. There are two main technical challenges facing SATDID, namely \emph{i) recommending when hidden instances of technical debt in code should be self-admitted}, and \emph{ii) generating SATD comments describing the hidden TD in the identified code fragments}. 
SATDID's design consists of multiple components that are distributed across four processing modules (see Figure \ref{fw_arch_des}) to address these challenges. The four processing modules are Data Processing, Data Vectorisation, \ac{satd} Identification, and \ac{satd} Comment Generation. The first two modules prepare for and facilitate facing the technical challenges while the last two modules are responsible for the direct handling with them. These framework components operate in a chronological manner to achieve the main objective of providing \ac{satd} recommendation and description services.


In the Data Processing module, the input source code passes through the component responsible for code processing (e.g. parsing a conditional statement and building its Abstract Syntax Tree). For the next two modules, the user can choose between two configurations: either using a traditional machine learning classifier (e.g. \ac{mnb}, \ac{svm}, and \ac{rf}) or deep learning classifier (e.g. \ac{rnn} and \ac{cnn}). There are two main reasons for choosing this configuration setup. The first reason is to study and report the differences between the performances of the two configurations. The second reason is to provide the ability to the end user to choose their preferred model based on their available data and machinery when using our framework. Section \ref{dl-vs-ml} is dedicated to further discuss this point. Each configuration has its own data vectorisation technique. In the Data Vectorisation module, the Word Embedding component creates an embedding (i.e. vector) for each token in the processed source code and passes it to the deep learning component in the next module. The \ac{vsm} component creates a vector that represents the entire processed source code and passes it to the traditional machine learning component in the next module. The SATD Identification module contains the machine/deep learning components responsible for identifying if there is \ac{td} in the code that should be self-admitted. If this is the case, SATDID flags it and passes its vector representation to the \ac{satd} Comment Generation module. The \ac{satd} Comment Generation module contains a deep learning component which generates the appropriate \ac{satd} comment that can be attached to the input code fragment. More details on our framework modules are presented in sections \ref{sect:process-vector}, \ref{satd-id}, and \ref{satd-com-gen}.
\section{Data Processing and Vectorisation}\label{sect:process-vector}

In this section, we present our implementation for the first two modules of SATDID, namely the Data Processing and Data Vectorisation modules. These modules are responsible for transforming an input code fragment into the appropriate format required by the machine learning components in the next modules. Sections \ref{data-process} and \ref{data-vector} describe the Data Processing and Data Vectorisation modules, respectively, in details. 


\subsection{Data Processing}\label{data-process}

The Data Processing module consists of a source code processing component and a comment processing component. Note that we process the SATD comments in the training data used for generating SATD comments. We do not use them for SATD identification since our approach caters for cases where those comments do not exist. Hence, we describe here the code processing component. We discuss the comment processing component in Section \ref{dataset-prep}, where we describe our pre-processing of the data used for training our models. 

\subsubsection{Source Code Processing}\label{sc-proc}


First, we create the \ac{ast} of the input source code fragment. Second, we create a sequential representation of the tree using a method proposed by Hu et al. \cite{hu2018deep} called \ac{ast} with \ac{sbt}. 
Classical traversal methods that convert \ac{ast}s to sequences (e.g. pre-order traversal) can be ambiguous in the way that different code fragments may produce the same sequence representation. We adapt the \ac{sbt} representation proposed in Hu et al. \cite{hu2018deep} to ensure unique sequence representations for different code inputs.

Suppose we have an \ac{ast} with only three nodes: a parent node and two child nodes. Let us call the parent `1', the left-child `2', and the right-child `3'. The \ac{sbt} method will represent the tree with the following sequence: \texttt{(1(2)2(3)3)1}.

\begin{sidewaysfigure}
    \centering
    \includegraphics[width=0.8\linewidth]{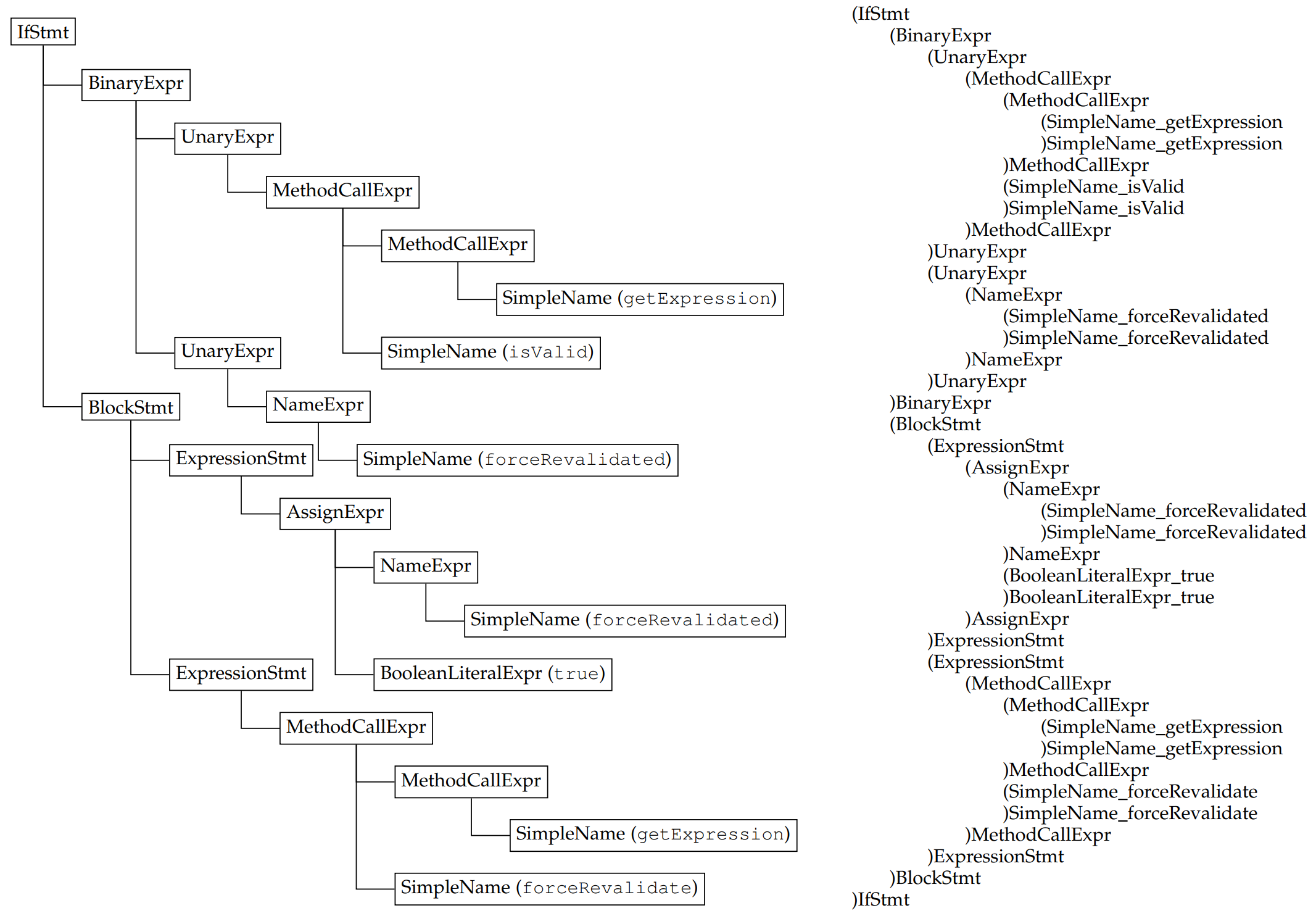}
    \caption{The Abstract Syntax Tree (AST) of the conditional statement in Scenario 2.b in Figure \ref{motiv_eg} is on the left-hand side. The sequence representation of the \ac{ast} using the Structure-Based Traversal (SBT) method is on the right-hand side.}
    \label{ast_with_sbt}
\end{sidewaysfigure}

In \ac{ast}s, non-terminal nodes represent the structural information of the code, and they have a ``type''. Terminal nodes have a ``type'' and a ``value'', where ``value'' is the concrete source code token and ``type'' is its type. In \ac{sbt}, non-terminal nodes are represented by their types, while terminal nodes are represented by their types and values. Figure \ref{ast_with_sbt} shows how the Source Code Processing component in the Data Processing module parses the conditional statement in Scenario 2.b in Figure \ref{motiv_eg}
to its \ac{ast} then applies the \ac{sbt} method to produce the sequence representation of the tree.

\subsection{Data Vectorisation}\label{data-vector}

The Data Vectorisation module receives the processed code fragment from the Data Processing module. The Data Vectorisation module is responsible for creating vector representations of code fragments, on/from witch the learning components of the framework train/predict. This module contains two components, namely the Word Embedding component (see Section \ref{word-embed}), which is linked to the deep learning component in the next module, and the Vector Space Model (VSM) component (see section \ref{bow-tfidf}), which is linked to the traditional machine learning component in the next module.

\subsubsection{Word Embedding}\label{word-embed}

Text data typically are highly sparse \cite{bingham2001random} and of high dimensionality \cite{aggarwal2012mining}. If we use one-hot encoding to create the vector representation of a word, we will end up with a vector that is the size of the entire vocabulary with all 0s except a 1 at the word's position, and that is for each word in the vocabulary of our dataset. Note that ``word'' here refers to either an \emph{AST token} (in source code) or a \emph{textual word} (in comments). To alleviate this problem, we use a technique called \emph{word embedding} \cite{gal2016theoretically} which aims to represent each word in the vocabulary as a fixed-length continuous vector (also called an embedding). The values in those embeddings are learnt and adjusted during model training. Word embedding has the trait of finding semantic relations between words/tokens according to how the values inside their embeddings relate to each other. Embeddings that have common semantics tend to cluster together.


Central to this is an embedding matrix
$\mathcal{M}\in\mathcal{R}^{d\times\lvert\mathcal{V}\rvert}$ where $\boldsymbol{d}$ is the embedding size and $\boldsymbol{\lvert\mathcal{V}\rvert}$ is the number of words in our vocabulary $\boldsymbol{\mathcal{V}}$. The embedding matrix $\boldsymbol{\mathcal{M}}$ acts as a lookup table where each row is a vector representation of a word in our vocabulary. Each word will have an index, and the word with index $\boldsymbol{i}$ will have its vector representation (i.e. embedding) at the $\boldsymbol{i^{th}}$ row of the matrix $\boldsymbol{\mathcal{M}}$. Machine/Deep learning models only deal with these indices and their vector representations and do not have access to the actual words/tokens. SATDID will generate indices that then can be converted to their associated words from our vocabulary in order to display the generated comment sentences.

\subsubsection{Vector Space Model}\label{bow-tfidf}


Vector Space Model (VSM) \cite{salton1975vector} is the compatible vectorisation technique with traditional machine learning. While every word/token is represented as a vector in word embedding (Section \ref{word-embed}), the entire source code fragment is represented by a vector in \ac{vsm}. Let us call a source code fragment a document $\boldsymbol{d}$, and a code token a term $\boldsymbol{t}$. Every document $\boldsymbol{d}$ is represented by a vector, every vector is a data point, and every term $\boldsymbol{t}$ is a dimension in these vectors. We calculate the weights of these terms by using a scheme called \ac{tfidf}. This scheme determines the ``importance'' of a term $\boldsymbol{t}$ in a document $\boldsymbol{d}$. Term's importance to a document increases by two factors: its frequency in that document and its rarity in the entire document set. \ac{tfidf} is computed as follows:

\begin{equation}
idf(t) = \log(\frac{\lvert\mathcal{D}\rvert}{df(t)}) + 1
\end{equation}

\begin{equation}
tfidf(t, d) = tf(t, d) \times idf(t)    
\end{equation}

where $\boldsymbol{\lvert\mathcal{D}\rvert}$ is the total number of documents, $\boldsymbol{df(t)}$ is the number of documents the term $\boldsymbol{t}$ appears in, and $\boldsymbol{tf(t, d)}$ is the number of times the term $\boldsymbol{t}$ appears in document $\boldsymbol{d}$.

\section{SATD Identification}\label{satd-id}

This module handles the vectors produced by the Data Vectorisation module (Section \ref{data-vector}). It contains the learning components for identifying source code fragments which contain \ac{td} that should be self-admitted. There are two configurations here: using either a deep learning (Section \ref{sect:dldetect}) or traditional machine learning (Section \ref{ml-detect}) component.
Later, we will evaluate and discuss the implications of these configurations in sections \ref{sect:evaluation} and \ref{sect:discuss}.

\subsection{Deep Learning Detector}\label{sect:dldetect}

\begin{figure}[htb]
	\includegraphics[width=\linewidth]{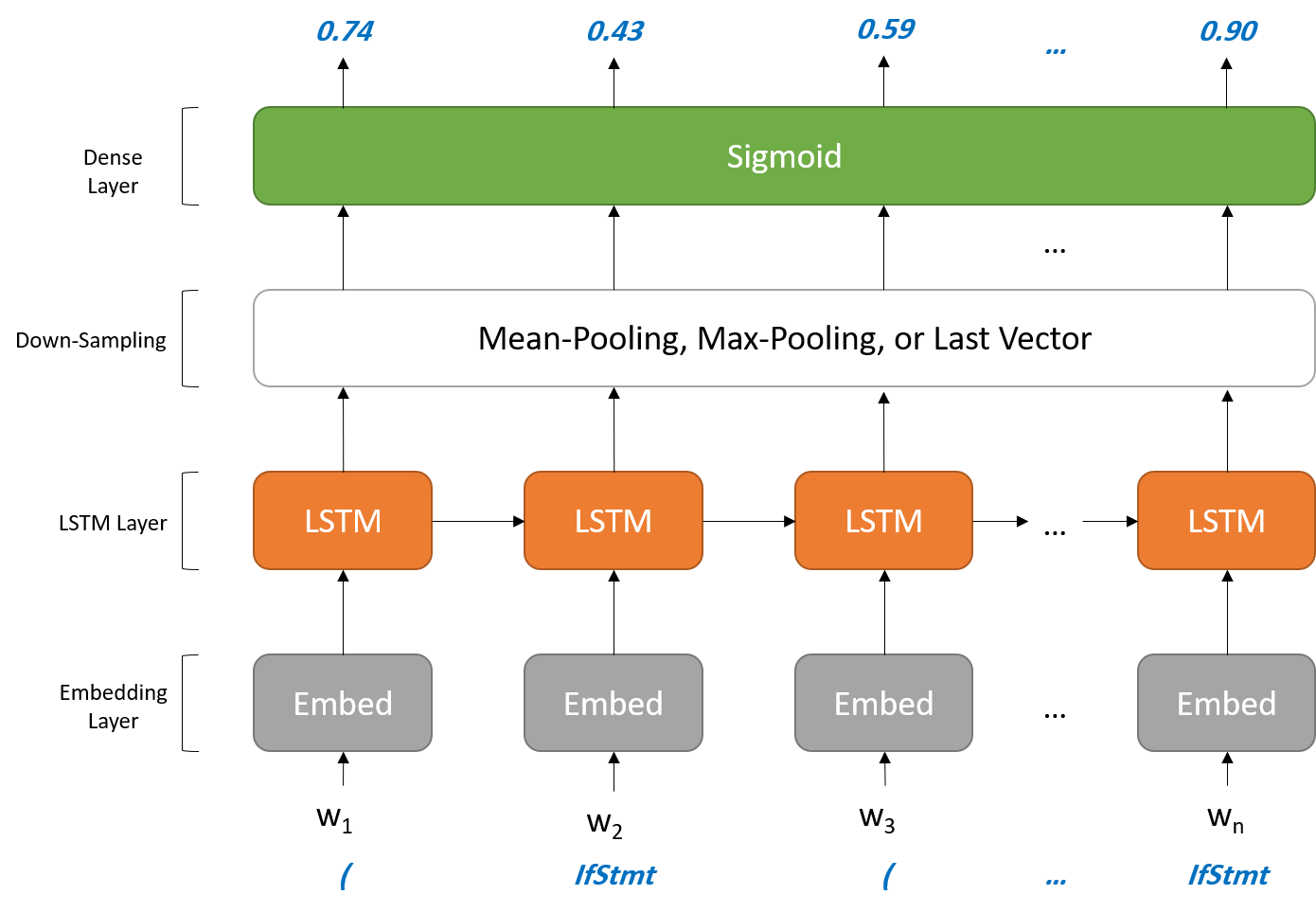}
	\caption{The structure of our deep learning detector (single-layered). The number of layers is determined by the number of stacked \ac{lstm} layers. An output closer to 1.0 recommends a \ac{satd} comment, while an output closer to 0.0 does not recommend a \ac{satd} comment.} 
	\label{class_struct}
\end{figure}

Our implementation of the deep learning detector (as well as the deep learning generator in Section \ref{satd-com-gen}) is based on the \ac{lstm} models \cite{hochreiter1997long}. Our deep learning detector has an embedding layer (see Section \ref{word-embed}) as its first layer. Let $\boldsymbol{W_1},...,\boldsymbol{W_n}$ be an input sequence produced by the programming language processing unit (see Section \ref{sc-proc}). The embedding layer converts the elements of the input sequence into their word embeddings (vectors) $\boldsymbol{V_1},...,\boldsymbol{V_n}$. The layer (or stack of layers) following the embedding layer is an \ac{lstm} layer (or a stack of \ac{lstm} layers, i.e. an \ac{lstm} network). An \ac{lstm} layer consists of a sequence of \ac{lstm} units. All of these units share the same model parameters since \ac{lstm}s are Recurrent Neural Networks (RNNs). At a time step $\boldsymbol{t}$, an LSTM unit reads an input vector $\boldsymbol{V_t}$ and the output state from the previous \ac{lstm} unit $\boldsymbol{S_{t-1}}$, and returns the current output state $\boldsymbol{S_t}$. Then, the output state $\boldsymbol{S_t}$ is passed onto two directions: the next layer (whether it is an \ac{lstm} layer or a Dense layer) and the \ac{lstm} unit in the next time step $\boldsymbol{t+1}$. The bottom two layers in Figure \ref{class_struct} depict the job of the embedding and \ac{lstm} layers.


There are several variations of our implementation of the deep learning detector which produce different results.
The variations include three different down-sampling techniques (Section \ref{down-sampl}) and different hyper-parameter settings (explained in Section \ref{exper-setup} and stated in Section \ref{eval-code-id}). 


\subsubsection{Pooling}\label{down-sampl}

While the deep learning model produces as many output state vectors as the number of LSTM units in the last layer of the LSTM network, the Dense layer accepts only one vector as an input. To down-sample the network's output to one vector, we examine the detector's performance with and without using a ``pooling'' technique. With pooling, we experiment with \emph{max-pooling} and \emph{mean-pooling}. Without pooling, we only consider the \emph{last} output state vector $\boldsymbol{S_t}$ at time step $\boldsymbol{t}$ since it holds information from all the previous output states $\boldsymbol{S_1...S_{t-1}}$, thanks to \ac{lstm} dynamics, and pass it to the Dense layer.
    
Pooling is a down-sampling method that reduces multiple inputs to the desired size (in our case, the size of the output state vector). At a time step $\boldsymbol{t}$, \emph{max-pooling} considers only the maximum values in every element position in the output state vectors $\boldsymbol{S_1...S_{t}}$, while \emph{mean-pooling} averages them. The resulting vector is then passed to the Dense layer.
    
Let us illustrate the three techniques using the following example. Suppose we have an input sequence of three items, $[W_1 \quad W_2 \quad W_3]$, and an output state of size 2. Suppose that the following are the final output states of these three items:
    
    
$\quad S_1 = S(W_1) = [5.2 \quad 3.3]$

$\quad S_2 = S(W_2) = [4.7 \quad 7.5]$

$\quad S_3 = S(W_3) = [9.1 \quad 0.6]$
    
We want to create one vector that captures information from all these three output states and use it as an input to the Dense layer. If no pooling technique is used, we consider the last output state $\boldsymbol{S_3}$. In \emph{max-pooling}, we pool the elements of the same position and consider only the maximum value. In \emph{mean-pooling}, we pool the elements of the same position and take the average:
    
$\quad NoPool(S_1, S_2, S_3)\;\;\;\;\,=[9.10 \quad 0.60]$
    
$\quad MaxPool(S_1, S_2, S_3)\;\:=[9.10 \quad 7.50]$
    
$\quad MeanPool(S_1, S_2, S_3)=[6.33 \quad 3.80]$

\subsubsection{Sigmoid Activation}\label{sig-activ}

As explained earlier, every \ac{lstm} unit is assigned for processing an input item, starting from the first item in the input sequence through the last item. At each time step $\boldsymbol{t}$, the vector resulting from down-sampling is passed to the Dense layer. The Dense layer has a \emph{sigmoid} activation function of the following formula:

\begin{equation}
S(x) = \frac{1}{1+e^{-x}}=\frac{e^x}{e^x+1}
\end{equation}

The sigmoid activation function returns a value between 0 and 1. This value represents what the detector ``thinks'' regarding the input sequence under investigation. If the value is closer to 1, it means that the detector leans towards deciding that there is hidden TD in the input code fragment which should be self-admitted. If the value is closer to 0, it means the detector votes for the opposite.

\subsection{Traditional Machine Learning Detector}\label{ml-detect}
An alternative implementation to deep learning is using traditional machine learning algorithms.
We feed the \ac{tfidf} vectors prepared by the previous module (see Section \ref{bow-tfidf}) to our traditional machine learning detector. Based on our experimentation with many machine learning algorithms, Support Vector Machines (SVM) and Multinomial Naive Bayes (MNB) provide the best comparable performances to deep learning for SATD identification. We also experimented with Random Forest (\ac{rf}) as part of replicating the benchmark's approach (see Section \ref{recom-bench}).

\ac{svm} is a machine learning algorithm that maximises the margin between the class-separating line, i.e. the hyperplane, and the closest data points of the dataset's classes \cite{joachims2001statistical}. \ac{mnb} implements the naive Bayes algorithm for multinomially distributed data \cite{rennie2003tackling}. \ac{rf} \cite{breiman2001random} is an ensemble of Decision Trees, where each tree depends on an independent random vector and all the trees share the same distribution. \ac{svm} is a leading approach for text categorisation problems as suggested by Kibriya et al. \cite{kibriya2004multinomial}. In addition, McCallum et al. \cite{mccallum1998comparison} argue that \ac{mnb} proves effectiveness with large vocabulary sizes. We refer the reader to \cite{kibriya2004multinomial,joachims1998text,xu2012improved} for further details on \ac{svm}, \ac{mnb}, and \ac{rf} for text classification.

\section{SATD Comment Generation}\label{satd-com-gen}

\begin{figure}[htb]
	\includegraphics[width=\linewidth]{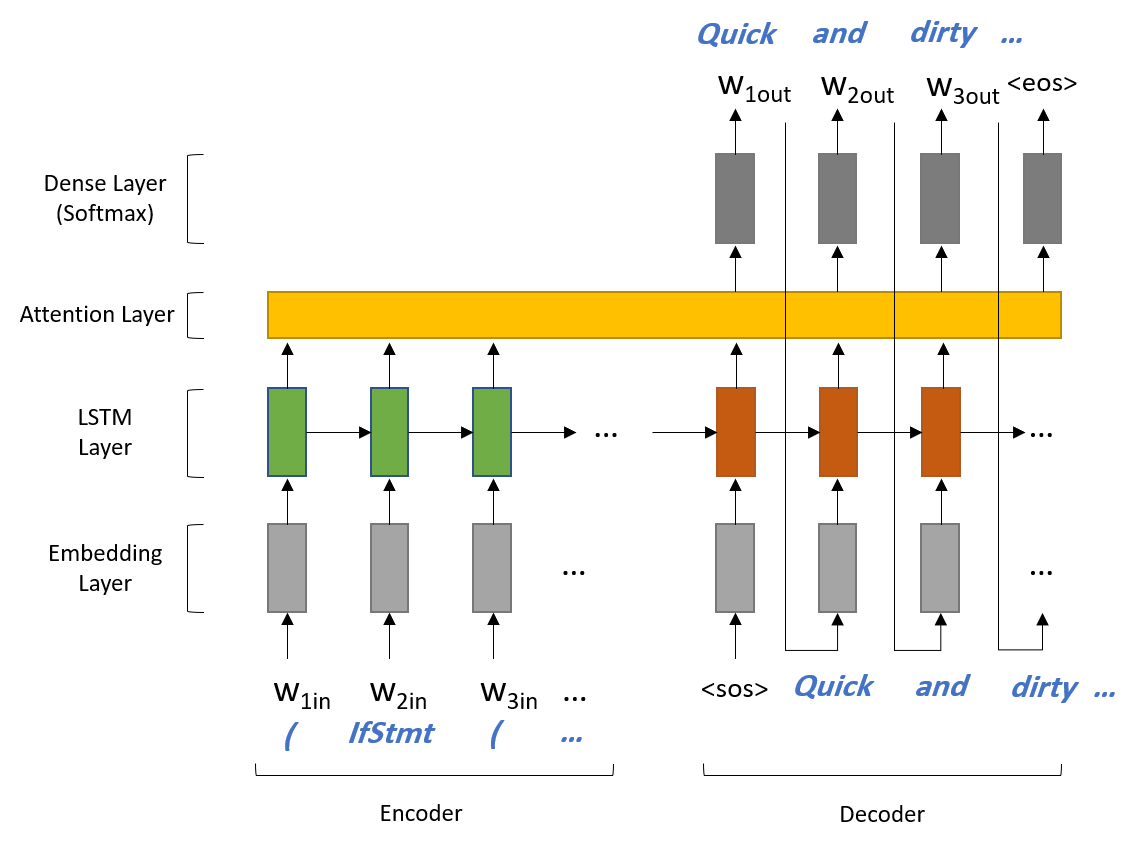}
	\caption{The structure of our deep learning generator (single-layered encoder and decoder). The number of layers is determined by the number of stacked \ac{lstm} layers.}
	\label{gen_struct}
\end{figure}

The purpose of the SATD Comment Generation module
is to generate an appropriate \ac{satd} comment that describes the \ac{td} in a code fragment. We implement the comment generator in this module using the deep learning \emph{encoder-decoder} model which employs the sequence-to-sequence (\emph{seq2seq}) learning method \cite{cho2014learning, sutskever2014sequence}. 
The encoder and decoder are two \ac{lstm}-based networks. We also incorporate the Attention mechanism (see Section \ref{att-mech}) and Beam search (see Section \ref{beam}) into our generator. We examine the generator's performance in multiple hyper-parameter settings (explained in Section \ref{eval-com-gen}). An internal view of our \ac{satd} comment generator is depicted in Figure \ref{gen_struct} (adapted from \cite{luong2015effective}).

The dynamics in which the embedding and \ac{lstm} layers of our comment generator operate is the same as the layers in our deep learning detector (described in Section \ref{sect:dldetect}). The difference between the generator and the detector (other than the generator being a composite of two \ac{lstm}-based networks) is highlighted in the top layers. In every time step $\boldsymbol{t}$, the generator passes the output state $\boldsymbol{S_t}$ to the Attention layer (see Section \ref{att-mech}) instead of a Dense layer. In addition, the encoder passes its last output state $\boldsymbol{S_n}$ to the first \ac{lstm} unit in the decoder to accompany the embedding of the pre-first target comment word $\boldsymbol{W_{out\_0}}$ (which we define as \texttt{<sos>}). The \ac{lstm} layer(s) in the decoder produces the first output state $\boldsymbol{S_1}$. $\boldsymbol{S_1}$ is passed to next \ac{lstm} unit as well as the Attention layer. The vector resulting from the Attention layer is passed to the Dense layer in order to predict the first target word $\boldsymbol{W_{out\_1}}$. For predicting every target word $\boldsymbol{W_{out\_t}}$, the decoder is fed with the previous target word $\boldsymbol{W_{out\_t-1}}$. This training technique is called teacher-forcing \cite{chollet2017kerasseq2seq}, where the decoder trains to generate the same target sequence but offset by one time step.

\subsection{Attention Mechanism}\label{att-mech}

The Attention Mechanism has demonstrated remarkable improvements in \ac{nmt} tasks \cite{bahdanau2014neural}. We implement an Attention layer into our \ac{satd} comment generator
to align between certain items in the input and output sequences. When predicting an output comment word $\boldsymbol{W_{out\_t}}$ at a time step $\boldsymbol{t}$, the Attention layer determines the amount of contribution each token in the input sequence $\boldsymbol{W_{in\_1}},...,\boldsymbol{W_{in\_n}}$ has on generating the current output word $\boldsymbol{W_{out\_t}}$. Without the Attention layer, all the input tokens $\boldsymbol{W_{in\_1}},...,\boldsymbol{W_{in\_n}}$ would have the same weight when predicting $\boldsymbol{W_{out\_t}}$, which is less practical since certain input tokens can map more closely than others to the current output comment word. The Attention layer adjusts the weight mappings of input-output sequences gradually during training.

\subsection{Softmax Activation}\label{soft-activ}

The last layer in the decoder is a Dense layer with a \emph{softmax} activation function:

\begin{equation}
\sigma(x)_i = \frac{e^{x_i}}{\sum_{v=1}^{\lvert\mathcal{V}\rvert} e^{x_v}}
\end{equation}

If we have $\lvert\mathcal{V}\rvert$ words in our target vocabulary (in this case, the comment vocabulary), the \emph{softmax} activation function gives a probability value between 0 and 1 to each word in the vocabulary for the prediction at the current time $\boldsymbol{t}$, where the sum of all these values is 1. The model then nominates the word with the highest probability value to be the predicted output word $\boldsymbol{W_{out\_t}}$ for the current position $\boldsymbol{t}$ in the comment sentence.

\subsection{Beam Search}\label{beam}

By default, the decoder uses greedy search to predict the likelihoods for the next word in the output sequence. Although this approach is often effective, it is non-optimal in some cases. In beam search, all the possible output words for the next step are generated, and the algorithm keeps track of the most likely $\boldsymbol{k}$ candidate sequences (in our case, comment sub-sentences up to the next output word). $\boldsymbol{k}$ is also known as the \emph{beam width}. Therefore, greedy search is a special case of beam search where $\boldsymbol{k} = 1$. Increasing the number of generated candidates $\boldsymbol{k}$ typically increases the possibility of finding the best candidate output sequence at the expense of a potential drastic decrease in decoder speed \cite{yoav2017neural,russell2002artificial,freitag2017beam}. We incorporate the beam algorithm during the comment generation process, which allows our model to generate multiple candidate comments for every code fragment that requires a \ac{satd} comment (more details in Section \ref{discuss-beam}).


\section{Model Training}\label{trn-dl}

The training data is fed to our deep learning components in \underline{batches}. For each batch, the neural network performs two training tasks: the \underline{feed-forward} task and then the \underline{back-propagation} task. In the feed-forward task, the model processes the input batch and calculates the predictions. In the back-propagation task, the model measures the error distance between the actual outputs (i.e. the ground truth) and the predicted outputs of the current batch, and tweaks its parameters accordingly. By performing the two training tasks, the model completes one \underline{training step}. Feeding the data in batches to the model has multiple benefits. Firstly, it accelerates the training process compared with feeding the model only one example at a time. Secondly, it introduces noise to the model which helps preventing over-fitting to the training data. Nonetheless, large batch sizes can be computationally exhaustive and reduce the prediction accuracy. \emph{Batch size} is one of the model's \emph{hyper-parameters} that we consider during hyper-parameter tuning (Section \ref{hp-tun}).

We apply an over-fitting prevention strategy called \underline{dropout} \cite{srivastava2014dropout}. At every training step, this strategy selects a random proportion of the neural network's nodes and stops it from processing the batch's examples. This is useful because some nodes tend to dominate the training weights. By setting a \underline{dropout rate} (we set ours to 20\%), we let the kept network nodes (80\%) process the current batch, which helps to avoid the weight dominance issue.

In the \ac{satd} Identification module, the deep learning detector's objective is to maximise the likelihood of predicting the target label (i.e. 1 if an input code fragment contains TD, and 0 otherwise). Suppose that the target label for a data point is $\boldsymbol{y}$ and the model prediction is $\boldsymbol{p}$. We already know the true value of $\boldsymbol{y}$ from the ground truth. The model uses this information to learn its weights. We measure the accuracy of our prediction $\boldsymbol{p}$ by calculating the \underline{log-loss} (also known as the \underline{cross entropy}) between $\boldsymbol{y}$ and $\boldsymbol{p}$:

\begin{equation}
-(y\log(p)+(1-y)\log(1-p))
\end{equation}

In the \ac{satd} Comment Generation module, the generator's objective is to maximise the likelihood of predicting the next target comment word. The target word at each time step is a word from our vocabulary $\mathcal{V}$. We can look at this situation as a \underline{multi-class classification problem}. Let $\boldsymbol{M}$ be the number of words in our target vocabulary ($M=\lvert\mathcal{V}\rvert$). We treat the problem as if we have $\boldsymbol{M}$ different classes. Suppose that at the current time step we are trying to predict the word \texttt{todo}, it is the fifth word in the vocabulary ($\boldsymbol{c}=5$), and we have only ten words in our vocabulary ($\boldsymbol{M}=10$). Since it is a multi-class classification problem, we will have 10 different binary indicators ($\boldsymbol{y_{c,o}}$) for the current class. Only if the observation $\boldsymbol{o}$ is the same as the actual class (in this case, \texttt{todo} for both $\boldsymbol{c}$ and $\boldsymbol{o}$), then $\boldsymbol{y_{c,o}}$ is 1. $\boldsymbol{y_{c,o}}$ is 0 for the remaining 9 binary indicators (e.g. $\boldsymbol{c}$ is \texttt{todo} and $\boldsymbol{o}$ is \texttt{hack}). Therefore, the cross entropy for the current word prediction is calculated as follows:

\begin{equation}
\sum_{c=1}^{M} y_{c,o}\log(p_{c,o})
\end{equation}


Once the cross entropy is computed, a \underline{model optimiser} is used to update the model parameters on the opposite direction of the gradient of the log-loss. We use the \emph{Adam} optimiser \cite{kingma2014adam} in our deep learning detector and the \emph{RMSprop} optimiser \cite{choetkiertikul2018deep} in our generator to obtain the best model weights (i.e. model parameters) possible.

\section{Evaluation}\label{sect:evaluation}

We implement SATDID using scikit-learn\footnote{\url{https://scikit-learn.org/stable/}}, a machine learning library in Python, and Keras\footnote{\url{https://keras.io/}}, a Python deep learning library running on top of TensorFlow \cite{tensorflow2015-whitepaper}, a machine learning platform. We also utilise JavaParser\footnote{\url{http://javaparser.org/}} to help build the \ac{ast}s of source code fragments. To contribute to the software engineering research community, we have made our source code, dataset, and results publicly accessible\footnote{\url{https://github.com/Abdulaziz-Alhefdhi/tech_debt}}.

We explain our dataset collection and pre-processing in Section \ref{sect:data_emp}. We describe our experimental setup in Section \ref{exper-setup}. The evaluation metrics used for our study are presented in Section \ref{eval_mtrcs}. We present the results of our experiments for \ac{satd} Identification and \ac{satd} Comment Generation in Sections \ref{eval-code-id} and \ref{eval-com-gen}, respectively.


\subsection{Dataset}\label{sect:data_emp}

In Section \ref{data-collect}, we explain the criteria and the procedures of our dataset collection. In Section \ref{dataset-prep}, we describe the steps taken to pre-process our dataset in order to have it in a framework-ready state.

\subsubsection{Data collection}\label{data-collect}

To train our framework, we need to prepare a dataset of code-comment pairs, where some of the pairs are \ac{satd} pairs and some of them are not.
Since conditional statements are said to be error-prone program elements~\cite{10.1007/s10664-013-9282-8,10.1109/TSE.2016.2560811,9193975}, this study focuses on conditional statement and comment pairs.

Code comments were collected with the same procedure in the previous study~\cite{id1288}, which had targeted active software development repositories on GitHub. We targeted repositories written in Java.
Active software development repositories were selected from the MySQL dabase dump 2018-04-01 of GHTorrent datasets~\cite{Gousios:2013:GDT:2487085.2487132} with the following criteria~\cite{id1288}: (i) more than 500 commits (the same threshold used in previous work~\cite{Aniche:2018:CSM:3238579.3238606}), and (ii) at least 100 commits in the most active two years (to remove long-term less active projects and short-term repositories, which may not be software development projects~\cite{Munaiah:2017:CGE:3147777.3147808}).

From the collected 4,995 Java repositories, 
single comments and the conditional statements immediately following them were collected as pairs of code and comment.
%
By analysing the \ac{ast} of each source file with an ANTLR4-based Java parser,
``outermost'' \texttt{if}-statements were identified.
We ignored inner conditional statements enclosed in another \texttt{if}-statement. 
A sequence of \texttt{else-if} (e.g. \texttt{if-else-if-else-if ...else}) is regarded as a single \texttt{if}-statement. An \texttt{if}-statement is linked to a comment if the comment satisfies the following two conditions: (i) It appears between the \texttt{if} keyword and its previous non-comment token, and (ii) The character position in line is the same as the \texttt{if}-statement. Although multiple comments may link to an \texttt{if}-statement, we removed them from our dataset.

From the extracted comments, we prepare \ac{satd} and non-\ac{satd} comments using the following keywords shown in the previous study~\cite{huang2018identifying}.
\begin{itemize}
    \item \ac{satd} comments: including at least one of the common 14 single keywords of \textit{todo}, \textit{fixme}, \textit{hack}, \textit{workaround}, \textit{yuck}, \textit{ugly}, \textit{stupid}, \textit{nuke}, \textit{kludge}, \textit{retarded}, \textit{barf}, \textit{crap}, \textit{silly}, and \textit{kaboom}.
    \item Non-\ac{satd} comments: excluding all the above 14 keywords and other frequently appearing 22 keywords of \textit{implement}, \textit{fix}, \textit{ineffici}, \textit{xxx}, \textit{broken}, \textit{ill}, \textit{should}, \textit{need}, \textit{here}, \textit{better}, \textit{why}, \textit{method}, \textit{could}, \textit{work}, \textit{probabl}, \textit{not}, \textit{move}, \textit{more}, \textit{make}, \textit{code}, \textit{but}, and \textit{author}.
\end{itemize}
We obtained 5,313 \ac{satd} code-comment pairs and 839,431 non-\ac{satd} code-comment pairs. In the collected 5,313 \ac{satd} pairs, there are 2,851 distinct comment contents. 



To understand the characteristics of the collected \ac{satd} code-comment pairs, 
a statistically representative sample of the distinct \ac{satd} comments was analysed. The required sample size was calculated so that the ratio of publication citations would generalise to all comments with a confidence level of 95\% and a confidence interval of 5, and we obtained a sample of 339 code-comment pairs\footnote{\url{https://www.surveysystem.com/sscalc.htm}}.

Three authors independently investigated the same 20 pairs to determine whether (i) the comment represents technical debt in code, and (ii) conditional statements are single or multiple\footnote{Kappa agreement was calculated using \url{http://justusrandolph.net/kappa/}.}. The Kappa agreement levels were (i) 0.90 and (ii) 0.88, which indicate ``almost perfect''~\cite{viera2005understanding}. Based on this encouraging result, the remaining data was then investigated by a single author. In the statistically representative sample of 339 code-comment pairs, we found (i) 298 (88\%) are actually \ac{satd} pairs. Within the 298 \ac{satd} pairs, 270 (91\%) code segments are single \texttt{if}-statements.
We consider this result promising for our experiments as we collected \ac{satd} code-comment pairs with a small amount of noise and that the obtained conditional statements were not too complex, which is beneficial for learning \ac{satd} code-comment patterns. 

\subsubsection{Data Pre-Processing}\label{dataset-prep}



To avoid \ac{oom} and data noise issues, we set the maximum lengths for input sequences and comment sentences to 1500 and 150 token/word, respectively. We do not truncate \ac{ast} sequences and comments. Truncation could be useful in accelerating classification tasks (e.g. \ac{satd} Identification). However, it could harm the \ac{satd} Comment Generation task since truncated words in the output sentence could map to tokens in the input sequence and vice versa. Thus, data points longer than the maximum lengths are ignored. We reserve an \texttt{<UNKN/PAD>} token for (i) padding during model training and, (ii) at model validation/testing, replacing input tokens that have not been seen during model training. 

For processing the comments, we ignore numbers, non-English text, and special characters. A start-of-sentence token, \texttt{<sos>}, is added to the beginning of every comment, and an end-of-sentence token, \texttt{<eos>}, is added to the end of every comment.
As can be seen in Figure \ref{gen_struct}, we need the start-of-sentence token to signal to the deep learning model to generate the actual first word in the comment, and we need the end-of-sentence token as a signal for the model to stop the comment generation process.

To ensure that there is no bias towards a subset of the dataset, and to avoid data leakage to the training set \cite{kaufman2012leakage}, we enforce a strict rule that removes all duplicate instances in the dataset so that every data point in the dataset is unique. Additionally, we apply a data randomisation procedure using the Mersenne Twister pseudorandom number generator \cite{matsumoto1998mersenne} to avoid order biases that may or may not have occurred during the time of data collection.

After performing the pre-processing steps, the number of \ac{satd} code-comment pairs shrinks from 5,313 to 3,022. Previous studies (e.g. \cite{potdar2014exploratory,zampetti2017recommending,huang2018identifying}) suggest that the percentage of SATD code in software projects ranges around 0.5-31\%. For the SATD Identification experiment, we have followed \cite{potdar2014exploratory} where they prove that the average percentage of SATD in software projects is 10.4\%. Therefore, we applied \emph{down-sampling} to the non-SATD class so that the ratio of SATD to non-SATD pairs in our dataset is around 1.4:8.6. 
We use both \ac{satd} and non-\ac{satd} pairs in order to teach the intelligent detectors to distinguish between the characteristics of \ac{satd} and non-\ac{satd} pairs. We use the rest of the non-\ac{satd} pairs (that were not used to train the detectors) in the \emph{pre-training} experiment (Section \ref{pre-train}). The dataset has 105,671 unique input tokens and 9,058 unique comment words.



\subsection{Experimental Setup}\label{exper-setup}

We perform and report the results of the 10-fold \ac{cv} of the intelligent components in our framework. For the deep learning components, we perform a Hyper-Parameter Tuning step first to find the appropriate hyper-parameter settings for the \ac{cv} step.

\subsubsection{Hyper-Parameter Tuning}\label{hp-tun}

This step is performed to search for the optimal set of model hyper-parameters. This step requires a separate validation set that will not be used during \ac{cv} for testing. We refer to this set as the ``tuning'' set instead of validation set to avoid confusion with cross validation. The tuning set is a stratified 10\% proportion of the entire dataset. In the hyper-parameter tuning step, we train our deep learning models on the remaining 90\% of the dataset multiple times while tuning the hyper-parameters every time. The best hyper-parameter settings are chosen according to the the models' performance results on the tuning set. The nominated hyper-parameter settings will then be used in the main \ac{cv} step (Section \ref{s10fcv}).

We experiment with a set of four hyper-parameters: the \emph{batch size}, \emph{number of layers}, \emph{layer size}, and \emph{embedding size}. The \emph{batch size} is discussed in Section \ref{trn-dl}. The \emph{number of layers} determines how many layers our \ac{lstm} network has. We experiment with one, two, and three layers. For the last two hyper-parameters (i.e. \emph{layer size} and \emph{embedding size}), we combine them in a super hyper-parameter and call it the \emph{latent dimensionality}. When we experiment with one and two layers, the size of the embedding and \ac{lstm} layers remain the same as the latent dimension. When we experiment with three layers, for the detector, the size of the embedding layer and the second \ac{lstm} layer are the same as the latent dimension, while the first \ac{lstm} layer is double the size of the latent dimension and the last \ac{lstm} layer is half. For the generator, all the layers are equal to the size of the latent dimension.

\subsubsection{10-Fold Cross Validation}\label{s10fcv}
We perform 10-fold cross validation (\ac{cv}) \cite{sklearn2019cv} on the entire dataset except the tuning set introduced in Section \ref{hp-tun}. By that, we guarantee every data point in the dataset is tested against. During \ac{cv}, we will use the tuning set for training but not testing. In other words, the tuning set will be included in the training set of each one of the folds in the 10-fold CV. This will give the model more observations to learn from as we put the tuning set to use instead of neglecting it. For the detectors, \ac{cv} is stratified. \ac{cv} is not stratified for the generator since that is not applicable. 10-Fold \ac{cv} is the main step whose performance will be evaluated and discussed next.

\subsection{Evaluation Metrics}\label{eval_mtrcs}

\subsubsection{Precision, Recall, and F-1 Scores}

We treat \ac{satd} Identification as a classification problems. Thus, we use \emph{Precision}, \emph{Recall}, and \emph{F1-Score} to evaluate our explored approaches and compare their performances against the benchmarks. \emph{Precision} indicates the rate in which the classifier is correct when claiming that a group of instances is a \ac{satd} group. \emph{Recall} indicates the rate in which the classifier is able to catch the \ac{satd} instances. Depending on the requirements of the project/situation, if practitioners do not care about identifying all \ac{satd} observations as much as the correctness of the identified ones, models with higher \emph{Precision} should be considered. On the other hand, if they are aiming at identifying as many \ac{satd} observations as possible and do not care as much about the correctness of the identified ones, models with higher \emph{Recall} should be adopted. Nevertheless, the \emph{F1-Score} is a measure that combines \emph{Precision} and \emph{Recall} together and is ideal for situations where the two metrics are equally important.

\subsubsection{Bleu-n Score}

We treat \ac{satd} comment generation as a translation problem. We use variations of the cumulative \emph{Bleu} score \cite{papineni2002bleu} to evaluate our approach. It has became a standard practice to use the \emph{Bleu} score to evaluate the performance of Neural Machine Translation (NMT).
\emph{Bleu} measures the similarity between the generated comments (the candidates) and the original comments from the ground truth in the dataset that were written by the developers (the references). \emph{Bleu} score produces a value between 0 and 1, inclusive, indicating how close the candidates are to the references, the higher the closer. For example, if a candidate is identical to a reference, the \emph{Bleu} score is 1. \emph{Bleu-n} calculates the cumulative similarity of \emph{n}-grams of text. For example, \emph{Bleu-4} calculates the similarity of \emph{1}-gram, \emph{2}-grams, \emph{3}-grams, and \emph{4}-grams, and then computes their weighted geometric mean. We report the results of \emph{Bleu-1}, \emph{Bleu-2}, \emph{Bleu-3}, and \emph{Bleu-4} in a percentage style. 

\subsubsection{Acceptability and Understandability}\label{accept-understand}

We also perform human evaluation between two authors on the comments generated by SATDID using two criteria, namely \emph{Acceptability} and \emph{Understandability}~\cite{oda2015learning}, to evaluate if the generated comments are easy to understand, especially for inexperienced programmers. We assigned a 5-level score (from 1 to 5) to indicate the acceptance of the generated comments, and a 6-level score (from 0 to 5) to show how well the annotators are in understanding of the generated comments.

\subsection{SATD Identification}\label{eval-code-id}



We experimented with the following hyper-parameter sets\footnote{We refer the reader to the full report if interested in the results of the hyper-parameter tuning step at \url{https://github.com/Abdulaziz-Alhefdhi/tech_debt}}:

\begin{itemize}
    \item \emph{Latent Dimensionality}: (8, 16, 32, 64, 128, 256)
    \item \emph{Number of Layers}: (1, 2, 3)
    \item \emph{Batch Size}: (8, 16, 32, 64, 128, 265, 512)
\end{itemize}

We also experimented with \emph{mean-pooling}, \emph{max-pooling}, and \emph{last-vector} (i.e. no-pooling). As pooling shows consistent performance improvements, we nominate the three best performing hyper-parameter settings with \emph{max-pooling} and \emph{mean-pooling} for the 10-fold \ac{cv} step. Table \ref{satd_cod_cv_res} lists the average \emph{Precisions}, \emph{Recalls}, and \emph{F1-Scores} of the 10-fold \ac{cv} step. The results are ordered according to the \emph{F1-Score}. 

Generally, adopting deep learning for this problem produces higher scores. Nonetheless, traditional machine learning algorithms provide comparable results.
For the deep learning detector, we can see that from the best six hyper-parameter settings presented in Table \ref{satd_cod_cv_res}, three of them has their \emph{Latent Dimensionality} size set to 32 and \emph{Batch Size} set to 256. Furthermore, none of the best six has 3 \ac{lstm} layers. 
For the traditional machine learning detector, \ac{mnb} provides a higher \emph{F1-Score} than \ac{svm}. However, \ac{svm} provides the highest \emph{Precision} score (41.5\%) amongst all the tested models. The highest Recall (29.8\%) and \emph{F1-score} (31.1\%) amongst all the tested models was achieved by the \ac{lstm} model with \{[64, 1, 64], \emph{max}\} for the \{[\emph{Latent Dimensionality}, \emph{Number of Layers}, \emph{Batch Size}], pooling technique\}. 


\begin{table}[htb]
	\centering
	\caption{Average \emph{Precisions} (P), \emph{Recalls} (R), and \emph{F1-Scores} (F1) of Stratified 10-fold Cross Validation of Our approach. The highest scores are in bold. LSTM: Long Short-Term Memory. TML: Traditional Machine Learning}
	\label{satd_cod_cv_res}
	\begin{tabular}{l|cccc|ccc}
		\toprule
		& \multicolumn{4}{c|}{\textbf{LSTM}} & \multirow{2}{*}{\textbf{P}} & \multirow{2}{*}{\textbf{R}} & \multirow{2}{*}{\textbf{F1}} \\
		& \textbf{\small Latent} & \textbf{\small Layers} & \textbf{\small Batch} & \textbf{\small Pool} \\
		\midrule
		1  & 64 & 1 & 64 & max & 34.2 & \textbf{29.8} & \textbf{31.1} \\
		2  & 32 & 1 & 256 & max & 34.6 & 27.9 & 30.5 \\
		3  & 32 & 2 & 256 & max & 35.0 & 25.2 & 28.7 \\
		4  & 32 & 2 & 256 & mean & 34.8 & 23.8 & 26.0 \\
		5 & 8 & 2 & 16 & mean & 40.1 & 24.5 & 25.7 \\
		6 & 16 & 2 & 64 & mean & 36.2 & 23.5 & 25.4 \\
		\midrule
        & \multicolumn{4}{c|}{\textbf{TML Algorithm}} & \textbf{P} & \textbf{R} & \textbf{F1} \\
		\midrule
		6 & \multicolumn{4}{c|}{Multinomial Naive Bayes} & 38.6 & 18.6 & 25.1 \\
		7 & \multicolumn{4}{c|}{Support Vector Machines} & \textbf{41.5} & 16.6 & 23.7 \\
		\bottomrule
	\end{tabular}
\end{table}

\subsubsection{Benchmarks}\label{recom-bench}

We benchmark against a machine learning tool (TEDIOuS) and a static analysis tool (SonarQube) for SATD Identification. We evaluate our approach by replicating/applying these benchmarks, run them on our dataset, and compare their results with SATDID's.

\textbf{TEDIOuS}: Zampetti et al. \cite{zampetti2017recommending} developed a Random-Forest-based approach called TEDIOuS. When a developer writes a new piece of code, TEDIOuS recommends to them if they should self-admit ``design'' technical debt. To the best of our knowledge, this is the only existing work in the \ac{satd} field that analyses the source code instead of the comment. Unlike our approach, they only focus on design debt, and they build the feature space using source code metrics instead of using the concrete source code. More details of Zampetti et al.'s approach can be found in Section \ref{sect:related-work}. We replicate their approach and use it as a benchmark. Table \ref{satd_cod_cv_compar} orders TEDIOuS's results alongside the other experiments based on \emph{F1-Score}, the highest first. 




\textbf{SonarQube}: \ac{sat} are prominently used as means to improve code quality by revealing recurrent code violations without incurring the costs of running the program \cite{marcilio2019static}. One of the most famous \ac{sat} is SonarQube\footnote{\url{https://www.sonarqube.org/}}. SonarQube is an automatic code review tool to detect bugs, vulnerabilities, and code smells in the code. We use SonarQube as another benchmark in order to compare our approach with \ac{sat} in recommending \ac{satd}.
To conduct this experiment, we leverage SonarQube's code smell analysis capability. Table \ref{satd_cod_cv_compar} provides a comparison of the results of using SonarQube alongside 
the other experiments.

Table \ref{satd_cod_cv_compar} shows the results of SATDID implemented using both deep learning and traditional machine learning. We also experimented with two pre-training styles and without pre-training (see section \ref{pre-train} for pre-training details). 
The highest \emph{F1-Score} (31.1\%) was achieved by our \ac{lstm} model with \{[64, 1, 64], \emph{max}\}. This provides 31.78\% and 475.93\% improvements over TEDIOuS and SonarQube, respectively. The highest \emph{Precision} score (41.5\%) was achieved by our \ac{svm} (see Table \ref{satd_cod_cv_res}) with 23.88\% and 21.35\% improvements over TEDIOuS and SonarQube. The highest \emph{Recall} score (29.8\%) was also achieved by our \ac{lstm} model with \{[64, 1, 64], \emph{max}\}. This provides 59.36\% and 1,046.15\% improvements over TEDIOuS and SonarQube.
Therefore, our approach outperforms the two benchmarks in all the evaluation mertics.
We attribute this conclusion to SATDID's efficient feature extraction (executed by the Data Vectorisation module) and learning capabilities (provided by the \ac{satd} Identification module). 


\subsubsection{Pre-Training}\label{pre-train}

The purpose of this experiment is to see if pre-training can help initialise enhanced model weights for the main training time instead of random weight initialisation. The negligible difference between the deep learning and traditional machine learning results shown in Table \ref{satd_cod_cv_res} further motivated us to attempt pre-training. 
We tried two pre-training methods: \emph{end2end} and \emph{embedding pre-training with traditional machine learning}.

In \emph{end2end}, we train an \ac{lstm} model on predicting the next token in the input sequence. This results in a pre-trained embedding and \ac{lstm} layers. When we train the model for \ac{satd} Identification, We use the pre-trained layers whose weights are not randomly initialised anymore to see if it provides improved reseults.

In \emph{embedding pre-training with traditional machine learning},
we also train an \ac{lstm} model on predicting the next token in the input sequence. However, when we train the model for \ac{satd} Identification, we only use the embedding layer of the pre-trained model. We extract the vector representation of each token from the pre-trained embedding layer. Then, for every input sequence in our dataset, we take the mean-pooling of the embeddings (i.e. vectors) of its tokens in order to represent it as one vector.
After that, the resulting vectors are
fed to a traditional machine learning model. We tried different traditional machine learners for this experiment and found that \ac{svm} is the best performing one.

Table \ref{satd_cod_cv_compar} orders the results with and without pre-training alongside the other experiments based on \emph{F1-Score}, the highest first. 
Between the two pre-training styles, \emph{end2end DLD} achieved higher \emph{F1-Score} (30.8\%) and \emph{Recall} score (29.3\%), while \emph{embeddings with TMLD} achieved higher \emph{Precision} score (34.1\%). Contrary to our expectation, pre-training did not provide improvement to SATDID's performance. In terms of \emph{F1-Score}, \emph{end2end DLD} and \emph{embeddings with TMLD} show -0.96\% and -4.78\% performance declines to our deep learning and TML detectors, respectively. However, the pre-trained models still outperform the benchmarks. \emph{end2end DLD} provides 30.51\% and 470.37\% improvements over TEDIOuS and SonarQube, respectively. \emph{embeddings with TMLD} provides 1.27\% and 342.59\% improvements over TEDIOuS and SonarQube.

\begin{table}[htb]
	\centering
	\caption{Average results of Stratified 10-fold Cross Validation in comparison with two pre-training methods and two benchmarks. The highest \emph{Precision}, \emph{Recall}, and \emph{F1} scores are in bold. DLD: Deep Learning Detector. TMLD: Traditional Machine Learning Detector.}
	\label{satd_cod_cv_compar}
	\begin{tabular}{l|c|c|ccc}
		\toprule
		& \multicolumn{2}{c|}{Model} & \textbf{P} & \textbf{R} & \textbf{F1} \\
		\midrule
		1 & \multirow{4}{*}{SATDID} & DLD & 34.2 & \textbf{29.8} & \textbf{31.1} \\
		2 & & Pre-Trained end2end DLD & 34.0 & 29.3 & 30.8 \\
		3 & & TMLD & \textbf{38.6} & 18.6 & 25.1 \\
		4 & & Pre-Trained Embeddings with TMLD & 34.1 & 18.6 & 23.9 \\
		\midrule
		5 & \multicolumn{2}{c|}{TEDIOuS} & 33.5 & 18.7 & 23.6 \\
		6 & \multicolumn{2}{c|}{SonarQube} & 34.2 & 02.6 & 05.4 \\
		\bottomrule
	\end{tabular}
\end{table}

\subsection{SATD Comment Generation}\label{eval-com-gen}

The experimental setup of our generator slightly differs from the detectors' due to two reasons. First, the time and space complexity of training the generator is much higher than that of the detectors. Second, the generator reports distinctive results every time we tune the hyper-parameters. 
We experiment with the following hyper-parameter sets:

\begin{itemize}
    \item \emph{Latent Dimensionality}: (512, 1024, 2048)
    \item \emph{Number of Layers}: (1, 2)
    \item \emph{Batch Size}: (32, 64)
\end{itemize}


\subsubsection{Ground-Truth Evaluation}\label{gt-eval}

Table \ref{satd_gen_tun_cv_res} lists the results from both the hyper-parameter tuning and 10-fold \ac{cv} steps, ordered according to the \emph{Bleu-4} score, the highest first. The highest \emph{Bleu-n} scores in the hyper-parameter tuning step was achieved by the \ac{lstm} model with [1024, 1, 64] for [\emph{Latent Dimensionality}, \emph{Number of Layers}, \emph{Batch Size}].
Therefore, this hyper-parameter setting was nominated for the 10-fold \ac{cv} step. 

During the hyper-parameter tuning step, we started by setting the \emph{Latent Dimensionality} to 512 and gradually increased it. The distinctive behaviour of the generator clearly showed us that setting the \emph{Latent Dimensionality} to 1024 produces higher \emph{Bleu-n} scores as 512 and 2048 decreased the scores. Increasing the \emph{Number of Layers} to 2 gives the lowest \emph{Bleu-n} scores, so we kept experimenting with 1. This suggests that sometimes it is not ideal to over-complicate the model as that may lead to over-fitting to the training set and suppress the model's ability to generalise. We experimented with 64 and 32 for the \emph{Batch Size} and found that 64 trains faster and produces higher scores. Increasing the \emph{Batch Size} more than 64 caused \ac{oom} issues.



Table \ref{satd_gen_coms} shows some \ac{satd} comments generated by SATDID in comparison with human-written \ac{satd} comments from the ground-truth. The first example shows a generated comment that is identical to the comment written by the human developer. The second example shows minor differences, while the third example shows a comment that is totally different from the human-written one.

\begin{table}[htb]
	\centering
	\caption{\emph{Bleu-n} scores of the hyper-parameter tuning step followed by the average \emph{Bleu-n} scores of the 10-fold cross validation step. The highest \emph{Bleu-n} scores are in bold. We also apply cross validation on the benchmark's approach (Hu et al.) and report the results.}
	\label{satd_gen_tun_cv_res}
	\begin{tabular}{l|ccc|cccc}
		\toprule
		& \textbf{\scriptsize Latent} & \textbf{\scriptsize Layers} & \textbf{\scriptsize Batch} & \textbf{B-1} & \textbf{B-2} & \textbf{B-3} & \textbf{B-4} \\
		\midrule
		& \multicolumn{7}{c}{\textbf{Hyper-Parameter Tuning}} \\
		\midrule
		1  & 1024 & 1 & 64 & \textbf{15.6} & \textbf{10.7} & \textbf{09.1} & \textbf{08.3} \\
		2  & 1024 & 1 & 32 & 14.5 & 10.3 & 08.8 & 08.2 \\
		3  & 2048 & 1 & 32 & 13.3 & 09.1 & 07.7 & 07.0 \\
		4  & 512  & 1 & 64 & 11.7 & 05.5 & 03.3 & 02.3 \\
		5  & 1024 & 2 & 32 & 10.0 & 01.1 & 00.3 & 00.0 \\
		\midrule
		& \multicolumn{7}{c}{\textbf{10-Fold \ac{cv}}} \\
		\midrule
		\scriptsize{SATDID} & 1024 & 1 & 64 & \textbf{18.1} & \textbf{14.3} & \textbf{13.0} & \textbf{12.3} \\
		\scriptsize{Hu et al.} & 512 & 2 & 100 & 09.6 & 04.3 & 02.6 & 01.8 \\
		\bottomrule
	\end{tabular}
\end{table}


\begin{table}
\centering
	\caption{Sample model-generated \ac{satd} comments compared with human-written comments from the ground-truth.}
	\label{satd_gen_coms}
	\begin{tabular}{p{0.465\linewidth}|p{0.465\linewidth}}
		\toprule
		\multicolumn{1}{c|}{\textbf{Java Code}} & \multicolumn{1}{c}{\textbf{Comment}} \\
		\midrule
		\multicolumn{2}{c}{\textit{Human-Written Comments}} \\
		\midrule
		{\fontsize{7}{8.4}\selectfont\texttt{\textcolor{BlueViolet}{if(}number \textcolor{BlueViolet}{==} \textcolor{OliveGreen}{0}\textcolor{BlueViolet}{) \{} PrismObject \textcolor{BlueViolet}{<}UserType\textcolor{BlueViolet}{>} jack\textcolor{BlueViolet}{...}}} & \textcolor{gray}{// todo e g check metadata} \\
		\multicolumn{1}{c|}{-----------------------------------------------------} & \multicolumn{1}{c}{-----------------------------------------------------} \\
		{\fontsize{7}{8.4}\selectfont\texttt{\textcolor{BlueViolet}{if(}StringUtilities\textcolor{BlueViolet}{.}isNumeric\textcolor{BlueViolet}{(}itemName\textcolor{BlueViolet}{)) \{...}}}
		 & \textcolor{gray}{// workaround issue user types in something like mallsell pail} \\
		\multicolumn{1}{c|}{-----------------------------------------------------} & \multicolumn{1}{c}{-----------------------------------------------------} \\
		{\fontsize{7}{8.4}\selectfont\texttt{\textcolor{BlueViolet}{if(!}empty\textcolor{BlueViolet}{)} mView\textcolor{BlueViolet}{.}getControl\textcolor{BlueViolet}{()...}}} & \textcolor{gray}{// workaround clearing the text area is not a key event} \\
		\midrule
		\multicolumn{2}{c}{\textit{SATDID-Generated Comments}} \\
		\midrule
		{\fontsize{7}{8.4}\selectfont\texttt{\textcolor{BlueViolet}{if(}number \textcolor{BlueViolet}{==} \textcolor{OliveGreen}{0}\textcolor{BlueViolet}{) \{} PrismObject \textcolor{BlueViolet}{<}UserType\textcolor{BlueViolet}{>} jack\textcolor{BlueViolet}{...}}} & \textcolor{gray}{// todo e g check metadata} \\
		\multicolumn{1}{c|}{-----------------------------------------------------} & \multicolumn{1}{c}{-----------------------------------------------------} \\
		{\fontsize{7}{8.4}\selectfont\texttt{\textcolor{BlueViolet}{if(}StringUtilities\textcolor{BlueViolet}{.}isNumeric\textcolor{BlueViolet}{(}itemName\textcolor{BlueViolet}{)) \{...}}} & \textcolor{gray}{// workaround issue user types in something like shop reprice pail} \\
		\multicolumn{1}{c|}{-----------------------------------------------------} & \multicolumn{1}{c}{-----------------------------------------------------} \\
		{\fontsize{7}{8.4}\selectfont\texttt{\textcolor{BlueViolet}{if(!}empty\textcolor{BlueViolet}{)} mView\textcolor{BlueViolet}{.}getControl\textcolor{BlueViolet}{()...}}} & \textcolor{gray}{// hack for empty arrays} \\
		\bottomrule
	\end{tabular}
\end{table}

\subsubsection{Generic Comment Generation}\label{comgen-bench}
We have also compared our approach against existing techniques for generating comments from code. One of the prominent approaches was proposed by Hu et al. \cite{hu2018deep}. Hence, we have replicated their approach and run it on our dataset in order to benchmark our approach against it. For a detailed explanation of the differences between our approach and \cite{hu2018deep}'s, refer to Sections \ref{deal-lg-sq}, \ref{deal-oov}, \ref{bm-model-hp}, and \ref{discuss-beam}.

Table \ref{satd_gen_tun_cv_res} shows the \emph{Bleu-n} scores of SATDID and the benchmark (Hu et al.). The improvements our framework provides over the benchmark can be clearly noticed, where SATDID provides 88.54\%, 232.56\%, 400\%, and 583.33\% improvements in terms of the \emph{Bleu-1}, \emph{Bleu-2}, \emph{Bleu-3}, and \emph{Bleu-4}, respectively.
We attribute the performance improvements in our framework to our focus on SATD code-comment pairs as well as our approach towards the following four aspects (discussed in detail in Section \ref{discuss-satd-comgen}): dealing with long sequences, dealing with \ac{oov} tokens, model hyper-parameters, and including beam search.

\begin{table}[htb]
    \centering
    \caption{Performance of human evaluation between two evaluators according to \emph{Acceptability} and \emph{Understandability}}
    \label{tab:human-eval}
    \begin{tabular}{l|c|c}
        \toprule
         & \textbf{Acceptability} & \textbf{Understandability}   \\
        \midrule
        Mean (annotator 1) & 3.154 & 3.202  \\
        Mean (annotator 2) & 3.101 & 3.142  \\
        Total mean & 3.128 & 3.172    \\
        Correlations & 0.791 & 0.783    \\
        \bottomrule
    \end{tabular}
\end{table}

\subsubsection{Human Evaluation}
\label{eval-human-eval}



The third and fourth authors evaluated 337 randomly selected samples independently. The result of the evaluation presented in Table \ref{tab:human-eval} shows that the comments generated by SATDID are relatively \emph{acceptable} and \emph{understandable} by human. This is indicated by the mean score in both \emph{Acceptability} and \emph{Understandability} at 3.128 and 3.172 respectively. The assessment results between individual evaluators are quite similar. It can be seen in high positive correlation values, that is, 0.791 for the \emph{Acceptability}, and 0.783 for the \emph{Understandability}.

\section{Discussion}\label{sect:discuss}

In this section, we discuss some implications and lessons learned based on the results from our evaluation of the \ac{satd} Identification (Section \ref{discuss-satd-id}) and \ac{satd} Comment Generation (Section \ref{discuss-satd-comgen}) experiments. We also discuss the threats to the validity of our approach (Section \ref{threats}).


\subsection{SATD Identification}\label{discuss-satd-id}


\subsubsection{Deep Learning Versus Traditional Machine Learning}\label{dl-vs-ml}


Our experiments show that the deep learning detector performed only slightly better than the traditional machine learning detector. We recommend using the traditional machine learning detector due to its significantly smaller training time (a few seconds compared to $\sim$4 hours for training the deep learning detector). If one still prefers adopting deep learning, we recommend the adoption of a pooling technique, especially \emph{max-pooling}, as it consistently improves the model’s performance.



\subsubsection{Pre-Training}\label{disc-pretrain}

While pre-training is commonly used in practice, we do not recommend it for \ac{satd} Identification. Pre-training is an alternative to random weight initialisation. However, we have found that, in our setting, pre-training did not help improve the model's performance despite the substantial amount of time it took to complete. We attribute this behaviour to the size of our problem and dataset. Our dataset is relatively small and focuses on \ac{satd} code-comment pairs. Therefore, our models are able to learn during training and do not need the extra layer of complexity proposed by pre-training.


\subsubsection{Comparison with the Benchmarks}\label{disc-bm}

A key difference between our approach and TEDIOuS in \ac{satd} Identification is the input format to the model. TEDIOuS uses source code metrics as input features. On the other hand, our approach vectorises the concrete of source code, hence is able to capture (through learning) its syntactic and semantic structures. The reported performance improvement of our framework over the benchmarks' suggests the efficiency of our model input format.

We also notice the limitations of using static analysis tools for \ac{satd} identification. Although SonarQube provides comparable (but not higher) \emph{Precision} in our experiments, it fails in the \emph{Recall} and \emph{F1} aspects. This is attributed to SonarQube's speciality in detecting ``code smells" while our framework is concerned with broader technical debt patterns.



\subsection{SATD Comment Generation}\label{discuss-satd-comgen}

Our study focuses on the TD-affected source code. Our model examines the code to determine when to associate a SATD comment. If a code fragment is determined, the model generates the appropriate SATD comment. 
However, generating a comment for every code fragment, whether it is TD-affected or not, is beyond our scope and beyond the scope of Self-Admitted Technical Debt for that matter.

Sections \ref{deal-lg-sq}, \ref{deal-oov}, \ref{bm-model-hp}, and \ref{discuss-beam} describe the differences between our approach and the benchmark's in comment generation. Section \ref{on-eval-comments} discusses our evaluation process of the generated SATD comments, and Section \ref{sample-comments} discusses three sample SATD comments generated by our model.


\subsubsection{Dealing with Long Sequences}\label{deal-lg-sq}

Long sequences causes memory issues. We ignore long sequences in our approach while Hu et al. truncate them. Truncated target words can map to non-truncated input tokens and vice versa. With truncation, the model is forced to only map non-truncated elements in the source and target sequence pairs. This can result in incorrect mappings. To avoid that, we ignore sequences that passes the thresholds instead of forcing incorrect mappings. When we replicated the benchmark, we increased the maximum lengths for codes and comments from 400 and 30 to 1500 and 150, respectively, to allow a chance for a better performance and to match with the maximum lengths of our approach. Despite that, the incorrect mappings still have their negative impact as shown in the reported results.

\subsubsection{Dealing with Out of Vocabulary Tokens}\label{deal-oov}

A large vocabulary size contributes to the time and space complexities of the problem. 
Hu et al. have a threshold for vocabulary size while we decided to tolerate the entire vocabulary. Not only did our approach 
allow us to tolerate the entire vocabulary, it also allowed the models to train faster. Thus, the time and space gains of the benchmark's approach of limiting the vocabulary size are not necessary for our approach, and we benefit from the increased accuracy of the vocabulary toleration. To replicate Hu et al.'s approach, only the most frequent 3.78\% source code tokens of our dataset will be included. Out of Vocabulary (\ac{oov}) tokens are represented by their types alone. For example, in Figure \ref{ast_with_sbt}, \texttt{choosing} is an \ac{oov} token, so it is represented by its type (\texttt{SimpleName}) alone, while \texttt{return} is represented by both its type (\texttt{ReturnStatement}) and value (\texttt{return}) since it is one of the most frequent 3.78\% tokens.

\subsubsection{Model Hyper-Parameters}\label{bm-model-hp}

Table \ref{satd_gen_tun_cv_res} illustrates the difference of our approach and the benchmark's in tuning three hyper-parameters. Hu et al.'s model hyper-parameters are [512, 2, 100] for [\emph{Latent Dimensionality}, \emph{Number of Layers}, \emph{Batch Size}], while ours are [1024, 1, 64]. A fourth hyper-parameter difference is the \emph{Number of Iterations} during training each fold of the 10-fold CV. We decreased it to 40 epochs (i.e. iterations) while the benchmark trains for 50 iterations. Our hyper-parameter choices for the 10-fold CV step were decided during the hyper-parameter tuning step. The reasoning behind our choices for \emph{Latent Dimensionality}, \emph{Number of Layers}, and \emph{Batch Size} are explained in Section \ref{gt-eval}. For the \emph{Number of Iterations} hyper-parameter, our experiments show that our model reaches convergence at 40 epochs at most. Yet, we achieved higher results with less iterations.

\subsubsection{Beam Search}\label{discuss-beam}

We added the beam algorithm to our model (see Section \ref{beam}) with $\boldsymbol{k} = 10$ which generates 10 candidate \ac{satd} comments for every input code fragment. Note that including the beam algorithm increased the training time from $\sim$14 hours (greedy search) to $\sim$26 hours, which is still comparable with the time spent by Hu et al.'s approach ($\sim$23 hours). Although the model's best candidate comment is usually the first candidate, beam search allows multiple attempts for the model to generate better comments when the first ones are not ideal. For example, one of the target comments in our dataset is:

\texttt{\textcolor{gray}{// hack to ensure the newly created swt fonts will be rendered with the same height as the awt one}}

Our model was able to generate the exact same comment in the fourth attempt (candidate 4), which had a positive impact on the \emph{Bleu-n} scores. 


\subsubsection{On the Evaluation of the Generated SATD Comments}\label{on-eval-comments}

One might assume that the \emph{Bleu-n} scores achieved by our generator are low. This is not necessarily the case as previous work (e.g. \cite{liu2018neural,wan2018improving}) reported similar and lower \emph{Bleu-n} scores. A limitation of \emph{Bleu-n} score is that it does not consider the semantic similarity. For example, if the model generates a synonym for a word, a human can accept that. However, this does not increase the \emph{Bleu-n} score. The human evaluation of the generated SATD comments in this work (Sections \ref{accept-understand} and \ref{eval-human-eval}) is conducted to address this issue.

\subsubsection{Samples from the Generated SATD Comments}\label{sample-comments}

We will discuss here three samples from the generated SATD comments and compare them to the SATD comments from the ground-truth in the dataset. Sample 1 shows a generated comment that is identical to the ground-truth comment. Sample 2 shows a comment that is completely different from the Ground-Truth comment, while Sample 3 shows a comment slightly different from the ground-truth comment. \\

\noindent\fbox{%
    \parbox{\textwidth}{%
        \begin{center}
            \textbf{\large Sample 1}
        \end{center}
        Ground-truth SATD comment:
        
        \texttt{\textcolor{gray}{// this ugly block should give the vuln the close date of the last scan close map that it has}}
        
        Generated SATD comment:
        
        \texttt{\textcolor{gray}{// this ugly block should give the vuln the close date of the last scan close map that it has}}
    }%
}\\

The model has not seen the input code before for this comment but it has seen a similar code fragment. The model learned the characteristics of the code fragment and its associated comment and, therefore, was successful in generating the perfect SATD comment for the input code. This promising result demonstrates the applicability of SATDID in generating useful comments. \\


\noindent\fbox{%
    \parbox{\textwidth}{%
        \begin{center}
            \textbf{\large Sample 2}
        \end{center}
        Ground-truth SATD comment:
        
        \texttt{\textcolor{gray}{// fixme temporary workaround for inconsistencies}}
        
        Generated SATD comment:
        
        \texttt{\textcolor{gray}{// todo add a common interface to the mc grammar for all these productions and remove this hack}}
    }%
} \\

Although it is hard to determine why the model specifically generated this comment, we attribute the failure in this case as well as a few similar instances to the fairly small size of our SATD code-comment pair dataset. Despite the small dataset, we have reported an overall promising performance of our generator. We believe that the generator's performance will improve once a dedicated effort is made to collect a large corpus of SATD code-comment pairs. Hence, we urge the community to consider that moving forward in this research area. \\


\noindent\fbox{%
    \parbox{\textwidth}{%
        \begin{center}
            \textbf{\large Sample 3}
        \end{center}
        Ground-truth SATD comment:
        
        \texttt{\textcolor{gray}{// todo kludge}}
        
        Generated SATD comment:
        
        \texttt{\textcolor{gray}{// xxx kludge}}
    }%
} \\

This is one of the cases where the Bleu score decreases regardless of the semantic acceptability of the generated comment. \texttt{\textcolor{gray}{XXX}} is a generic tag used to replace other tags such as \texttt{\textcolor{gray}{TODO}} and \texttt{\textcolor{gray}{FIXME}}. Our model successfully generated a descriptive comment in this sample. With a larger dataset and longer training, we expect that the model will be capable of generating specific tags for different code fragments.

\subsection{Threats to Validity}\label{threats}


Threats to internal validity relate to biases in data labelling as well as the implementation and training of the framework components. Regarding data labelling, a threat to validity relates to the potential existence of actual SATD comments that do not pass our inclusion test and, thus, labelled non-SATD. To mitigate this threat, we conducted an investigation of the collected dataset (Section \ref{data-collect}). This investigation ensured that the \ac{satd} comments represent technical debt in source code. 
Regarding the implementation and training of the framework components, we have performed extensive hyper-parameter tuning experiments for all of the framework components. In the \ac{satd} Identification experiments, we also have tested multiple hyper-parameter settings in the cross validation step instead of only selecting the best performing setting from the hyper-parameter tuning step.


Threats to external validity relate to the quality and quantity of our dataset as well as our human evaluation procedure. Regarding the dataset, we have collected code-comment pairs from 4,995 active software development repositories. From these repositories, we obtained 5,313 \ac{satd} code-comment pairs and 839,431 non-\ac{satd} pairs.
Thus, we claim that these threats are minimal. Regarding the human evaluation procedure, we acknowledge the potential bias in having it conducted by the third and fourth authors. 

Since all the data is collected from open-source projects and repositories, there is an open question about the generalisability of our approach to company projects. However, developers in companies are less prone to self-admitting technical debt they introduce \cite{huang2018identifying,ren2019neural}, which lessens the threat to our approach's generalisability. Furthermore, all of the analysed projects and repositories are implemented in Java. Future work would involve exploring our approach with larger and more diverse programming languages and datasets.



\section{Related Work}\label{sect:related-work}


Potdar and Shihab \cite{potdar2014exploratory} proposed the concept Self-Admitted Technical Debt (SATD) to describe technical debt that is intentionally introduced and documented in the comments that accompany source code. They manually inspected 100k comments and extracted 62 patterns that are actively used in following studies to detect \ac{satd}. Research on Self-Admitted Technical Debt ever since has expanded \cite{maldonado2015detecting,wehaibi2016examining,bavota2016large,de2015contextualized,de2016investigating,maldonado2017empirical,zampetti2017recommending,zampetti2018self,yan2018automating,da2017using,huang2018identifying,ren2019neural}. Maldonado and Shihab \cite{maldonado2015detecting} examined 33k source code comments and determined five types of \ac{satd}: design debt, defect debt, documentation debt, requirement debt, and test debt. They found that design debt is the most common one. Farias et al. \cite{de2015contextualized,de2016investigating}
developed a \ac{cvmtd} to identify \ac{satd} in code comments by analysing code tags and words' parts of speech.


Zampetti et al. \cite{zampetti2017recommending} developed a Random-Forest-based approach called TEDIOuS. TEDIOuS recommends to developers when they should self-admit design debt. Zampetti et al. use static analysis tools to calculate method-level metrics in order to create the model input features. These metrics are structural metrics, readability metrics, and warnings raised by Static Analysis Tools. SATDID is inspired by their Zampetti et al.'s work. We use their approach as a benchmark for our \ac{satd} Identification components.
However, there are key differences between Zampetti et al.'s approach and ours. Firstly, they calculate source code metrics and use them as features for the machine learning model to learn from. In our case, we vectorise the source code tokens themselves rather than calculating metrics to create the feature space. Secondly, their work specialises in detecting ``design'' debt. Ours attempts to detect all kinds of technical debt. Thirdly, Their work operates on method-level. Ours can operate on any level (conditional-statement-level, method-level, class-level, etc.) as long as an Abstract Syntax Tree can be parsed from the target source code fragment. Lastly, We propose a comprehensive framework that also performs \ac{satd} Comment Generation and leverages deep learning as well as traditional machine learning.

Maldonado et al. \cite{da2017using} propose a \ac{nlp} approach to identify the two most common types of \ac{satd}: design debt and requirement debt. They employ a maximum entropy classifier to identify the \ac{satd} comments. Huang et al. \cite{huang2018identifying} used text mining techniques to detect all types of \ac{satd} comments. They utilise feature selection to select useful features for classifier training. Their work outperformed the \ac{nlp} approach. Note that a limitation of Maldonado et al.'s work is its speciality in only two \ac{satd} types rather than all types.
In their recent work, Ren et al. \cite{ren2019neural} Proposed a deep learning-based approach in the form of Convolutional Neural Networks (CNN) for \ac{satd} comment identification. They argue that leveraging CNN improves the detection task over other existing approaches. Furthermore, exploiting the computational structure of CNN improves the explainability of the results by identifying key patterns of \ac{satd} comments. 
The main difference between these approaches and ours is that our approach deals with the case when the comment is not provided with the source code, which is a key limitation of the previous work.
SATDID detects source code fragments that require \ac{satd} comments and, on top of that, generates the appropriate \ac{satd} comments for them, another contribution that is missing from the previous work.

\section{Conclusions and Future Work} \label{sect:conclusion}



In this paper, we have presented our framework, SATDID, for Self-Admitted Technical Debt recommendation and comment generation. To the best of our knowledge, we are the \emph{first} to propose such a comprehensive solution. SATDID analyses source code to determine if it contains technical debt that should be self-admitted. Our approach does not assume the existence of SATD comments as in existing work. In addition, if TD is found in a code fragment, the SATDID will generate an appropriate \ac{satd} comment which describes the TD and can be attached with the code fragment. For replicabilitiy purposes and further research in this area, we have made our code, dataset, and result reports publicly available. 

We have evaluated SATDID's performance on a dataset of code-comment pairs from 4,995 active software development repositories. 
Our approach provides at least 21.35\%, 59.36\%, 31.78\%, and 583.33\% improvements over all the replicated/tested benchmarks in terms of \emph{Precision}, \emph{Recall}, \emph{F-1}, and \emph{Bleu-4} scores, respectively. In addition, our approach scores total means of 3.128, and 3.172 in terms of \emph{Acceptability} and \emph{Understandability} of the generated comments, respectively. The results demonstrate the effectiveness of our approach.

Future work involves investigating larger and more diverse datasets. This includes studying other programming languages, types of code fragments other than conditional statements (e.g. method calls), as well as commercial and closed-source software projects when accessible. In addition, we plan to develop an automated, open-source plugin for \ac{satd} Identification and Comment Generation for Integrated Development Environments (IDEs). Furthermore, we plan to adopt other deep learning techniques in order to further validate its usage over traditional machine learning. Future work will also involve inviting external entities to extensively exam the human evaluation of the SATD comments generated by our model. Finally, we plan to extend this work and study \ac{satd} repayment (i.e. removal).

\bibliography{sn-bibliography}


\end{document}